\documentclass[12pt]{article}
\usepackage{amssymb,amsmath,epsfig}
\allowdisplaybreaks

\begin{document}
\title{\bf Scalar Field Cosmology via Noether Symmetries in Energy-Momentum Squared Gravity}
\author{M. Sharif \thanks {msharif.math@pu.edu.pk} and M. Zeeshan Gul\thanks{mzeeshangul.math@gmail.com}\\
Department of Mathematics and Statistics, The University of Lahore,\\
1-KM Defence Road Lahore, Pakistan.}

\date{}
\maketitle

\begin{abstract}
This article examines the analytic solutions of isotropic spacetime
with the minimal coupling of scalar field and matter in the context
of energy-momentum squared gravity. The scalar field includes
quintessence and phantom dark energy models. We evaluate solutions
of a physical system through the Noether symmetry approach that
provide some viable constraints to choose the cosmological models on
the basis of current observations. We consider different models of
this theory to establish Noether equations, symmetry generators and
corresponding conserved quantities. We compute analytic solutions
and examine their graphical representation for various cosmological
quantities. The behavior of these quantities is found to be
consistent with current observations implying that this theory shows
expanding behavior of the universe. We find that symmetry generators
and their conserved parameters exist in all cases.
\end{abstract}

solutions; Energy-momentum squared gravity.

\section{Introduction}

The current cosmic expansion has been the most fascinating
development for the researchers in the recent years. This cosmic
expansion is the result of a cryptic force dubbed as dark energy
(DE) which has repulsive behavior. Several researchers have been
working to expose its ambiguous characteristics. Cosmological
constant $(\Lambda)$ is the first candidate which determines the
ambiguous characteristics of DE, but it has \emph{fine-tuning} and
\emph{coincidence} problems. To address these issues, different
modified gravitational theories have been developed which are
considered the key factors to unveil dark aspects of the cosmos. The
$f(\mathbb{R})$ gravity is the simplest modified theory which has
significant literature \cite{1} to comprehend its physical features.
The concept of curvature-matter coupling generalized this modified
theory. These are non-conserved theories that confirm the presence
of an extra force, consequently, the path of the particle is
changed. These modified approaches are quite useful to understand
the cosmic mysteries. The minimal coupling is known as
$f(\mathbb{R},\mathcal{T})$ theory \cite{2} while the non-minimal
coupled theory is
$f(\mathbb{R},\mathcal{T},\mathbb{R}_{\xi\gamma}\mathcal{T}^{\xi\gamma})$
gravity \cite{3} One such coupling yields
$f(\mathcal{G},\mathcal{T})$ theory, where $\mathcal{G}$ is a
Gauss-Bonnet term \cite{3a}.

The beginning of the cosmos presents an interesting problem for
cosmologists. In this perspective, different researchers have tried
to formulate adequate cosmic models that determine cosmic evolution
since the advent of time. According to the big-bang theory, matter
and energy of the cosmos were initially concentrated at a single
point (singularity) having infinite density and temperature. The
universe came into existence due to the explosion of superheated
ultra-dense matter within the singularity and has been expanding
since then. Big bounce theory is an alternative technique that
demonstrates the evolution of the cosmos as a series of big bounces
(expanding and contracting) having no beginning or end state.
Singularities in general relativity (GR) are the most complicated
issue because of their prediction at a large scale, where GR is not
capable due to quantum effects. However, there is no particular
method in quantum gravity. Accordingly, energy-momentum squared
gravity (EMSG) is considered the most significant and elegant
technique that resolves the big-bang singularity in the non-quantum
description.

This modification of GR is developed by including a non-linear term
$(\mathbf{T}^{2}=\mathcal{T}_{\xi\gamma}\mathcal{T}^{\xi\gamma})$ in
the functional action also known as $f(\mathbb{R},\mathbf{T}^{2})$
theory \cite{4}. This newly developed theory establishes a
particular connection between geometry and matter as well as
includes an additional force that yields a more comprehensive
description to unveil the cosmic mysteries. This proposal contains
squared and product terms of matter contents which are useful to
examine different fascinating cosmological results. The main concept
behind EMSG is to resolve the big-bang singularity. In the early
times, this theory has a minimum scale factor and maximum energy
density which implies that there is a bounce in the early universe.
However, the cosmological constant does not play an important role
in the background of the standard cosmological model whereas the
antigravitational behavior of the cosmological constant resolves the
singularity only after the matter-dominated era in the EMSG. The
density profile supports the inflationary cosmological models which
sort out important cosmological problems. It is noteworthy that this
approach represents the entire cosmic history as well as the
evolutionary picture of the universe.

Roshan and Shojai \cite{5} analyzed that EMSG has a bounce at the
early universe and avoids the big-bang singularity. Board and Barrow
\cite{6} discussed the behavior of analytic solutions via cosmic
evolution, existence and absence of singularities in this theory.
The physically realistic and stable stellar objects through a
polytropic equation of state (EoS) have been studied in \cite{7}.
Bahamonde \cite{8} studied various EMSG models and observed that
these models determine the current cosmic evolution as well as
acceleration. The thermodynamic aspects of the compact objects and
stellar structures in the presence of quark matter have been
explored in \cite{9}. Recently, we have analyzed the stability of
the closed Einstein universe for specific EMSG models and found more
stable regions than GR and other alternative gravitational theories
\cite{10}. Also, the dynamics of self-gravitating objects in this
background have been examined \cite{11} and concluded that EMSG
reduces the collapse rate as compared to GR.

Symmetry determines the characteristics of a mathematical as well as
a physical system that remains invariant in the presence of some
change. The symmetry methods have played a crucial role in finding
the analytic solutions of non-linear differential equations. At the
geometric level, symmetry appears when a system remains preserved
under specific transformations such as rotation, reflection or
scaling. The symmetry which occurs as a result of constant change in
a system is known as continuous symmetry and the continuous symmetry
corresponding to the Lagrangian is called Noether symmetry (NS). The
viable attributes of a physical system can be determined by
formulating the corresponding Lagrangian which demonstrates the
energy content and provides information about possible symmetries of
the system. However, NS methodology is the most effective technique
that determines a relation between symmetry generators and conserved
values of the system \cite{12}. This strategy minimizes the system's
complexity and yields new solutions which can be used to discuss the
mysterious universe.

The NS technique is described in different ways in the literature
\cite{13}, i.e., one can obtain symmetry generators by setting the
Lie derivative of the Lagrangian equal to zero and another method to
find the generators is the addition of gauge term in the invariance
condition known as Noether gauge symmetry. The NS methodology is
important as it recovers some conservation laws and symmetry
generators of spacetimes \cite{14}. This approach is not just a
technique to deal with the dynamical solutions but their existence
also yields some valid constraints so that one can choose
cosmological models according to current observations \cite{15}.
Moreover, this methodology is a crucial and valuable tool for
determining exact solutions by employing conserved quantities.
Conservation laws are the key aspects to examine different physical
processes. These are the particular cases of the Noether theorem
which describes that there is some conservation law that corresponds
to each differentiable symmetry. The translation and rotation
symmetry of any object is determined by the conservation principles
of linear and angular momentum.

Noether charges are significant in the literature because they are
used to investigate various cosmological issues in different
considerations \cite{16}. Capozziello and Ritis \cite{17}
investigated exact cosmological solutions through the NS approach in
non-minimally coupled theory of gravity. Motavali and Golshani
\cite{18} determined some aspects of the coupling function for FRW
universe model through the NS approach. Vakili \cite{19} used this
technique and analyzed the current cosmic condition through EoS
parameter. Capozziello et al \cite{20} studied this technique in the
phantom as well as quintessence universe models. Sharif and Waheed
\cite{21} studied the Lie symmetry and NS approach to discuss the
structure of some black holes. Capozziello and Laurentis \cite{22}
examined the NS technique in the framework of quantum cosmology. Ali
and Hussain \cite{23} investigated positive energy bounds via NS
technique in Bianchi type-V universe model, whereas this work for
Bianchi type-I spacetime has been studied in \cite{24}.

Roshan and Shojai \cite{25} explored NS approach for
matter-dominated cosmos in Palatini $f(\mathbb{R})$ theory.
Capozziello et al \cite{26} obtained analytic solutions of the
sphere through NS strategy in $f(\mathbb{R})$ theory. Hussain et al
\cite{27} studied Noether gauge symmetry in the same theory. Shamir
et al \cite{28} found exact cosmological solutions of spherically
symmetric and FRW spacetimes through NS approach in $f(\mathbb{R})$
gravity. Kucukakca and Camci \cite{29} explored Noether gauge
symmetry of FRW spacetime in Palatini $f(\mathbb{R})$ theory. Jamil
et al \cite{30} studied scalar field cosmology through this
technique in teleparallel theory. Kucukakca \cite{31} found exact
solutions for a flat FRW universe in the scalar-tensor theory.
Momeni et al \cite{32} discussed exact cosmological solutions of
flat FRW spacetime through NS in $f(\mathbb{R},\mathcal{T})$ theory.
Sharif and Fatima \cite{33} studied this work in $f(\mathcal{G})$
gravity. Shamir and Ahmad \cite{34} used the same technique and
discussed the isotropic as well as anisotropic solutions in
$f(\mathcal{G},\mathcal{T})$ theory. Bahamonde et al \cite{35} found
new exact spherical solutions via NS in $f(\mathbb{R},\varphi,\chi)$
theory. Bahamonde et al \cite{36} discussed wormhole solutions
through the NS approach in $f(\mathrm{T})$ theory. Recently, we have
studied NS in $f(\mathbb{R},\mathbf{T}^{2})$ gravity \cite{37} and
also examined geometry of celestial objects and wormhole in this
framework \cite{38}.

The scalar field models are the key factors to examine the evolution
and the present state of the universe. In this perspective, various
canonical, i.e., \emph{phantom}, \emph{quintessence},
\emph{k-essence}, etc, and non-canonical models of scalar field have
been studied in \cite{39}. The non-canonical models are obtained
through the kinetic part of the scalar field that determines
late-time cosmic acceleration. Jamil et al \cite{40} established an
explicit form of the potential function for phantom and quintessence
phases through the NS approach in teleparallel gravity. Sharif and
Nawazish \cite{41} investigated scalar field cosmology with
different cosmic models through NS technique in
$f(\mathbb{R},\mathcal{T})$ theory. Fazlollahi \cite{42} used the NS
technique to formulate the analytic solutions of FRW spacetime and
examined their behavior through different cosmological quantities in
$f(\mathbb{R})$ theory.

This paper examines the NS approach for flat FRW universe admitting
scalar field models in the framework of EMSG. We find symmetry
generators with associated conserved quantities and obtain analytic
solutions to analyze the cosmic evolution. The paper is organized as
follows. Section \textbf{2} formulates the field equations and
point-like lagrangian of EMSG with the minimal coupling of matter
and scalar field. Section \textbf{3} gives a detailed study of the
NS technique. We establish exact viable solutions using the NS
approach and analyze their behavior graphically in section
\textbf{5}. We summarize our consequences in section \textbf{5}.

\section{Field Equations and Point-Like Lagrangian}

This section formulates the field equations and point-like
Lagrangian of the homogeneous and isotropic spacetime with a scalar
field in EMSG. The corresponding action is defined as \cite{4}
\begin{equation}\label{1}
\mathcal{S}=\int \sqrt{-\mathrm{g}}\big[\frac{1}{2\kappa^2}f
(\mathbb{R},\mathbf{T}^{2})+\mathcal{L} _{\mathrm{M}}\big]d^4x,
\end{equation}
where coupling constant and determinant of the metric tensor are
denoted by $\mathrm{g}$ and $\kappa^2$, respectively. The
corresponding field equations are obtained as
\begin{equation}\label{2}
\mathbb{R}_{\xi\gamma}f_{\mathbb{R}}+\mathrm{g}_{\xi\gamma}\Box
f_{\mathbb{R}}-\nabla_{\xi}\nabla_{\gamma}f_{\mathbb{R}}-\frac
{1}{2}\mathrm{g}_{\xi\gamma}f=\mathcal{T}_{\xi\gamma}-\Theta_{
\xi\gamma}f_{\mathbf{T}^{2}},
\end{equation}
where $\Box=\nabla_{\xi}\nabla^{\xi}$, $f\equiv f(\mathbb{R},
\mathbf{T}^{2})$, $f_{\mathbf{T}^{2}}=\frac{\partial f}{\partial
\mathbf{T}^{2}}$, $f_{\mathbb{R}}= \frac{\partial f}{\partial
\mathbb{R}}$ and
\begin{eqnarray}\nonumber
\Theta_{\xi\gamma}=-2\mathcal{L}_{\mathrm{M}}\left(\mathcal{T}_{
\xi\gamma}-\frac{1}{2}\mathrm{g}_{\xi\gamma}\mathcal{T}\right)-4
\frac{\partial^{2}\mathcal{L}_{\mathrm{M}}}{\partial\mathrm{g}^{
\xi\gamma}\partial\mathrm{g}^{\alpha\beta}}\mathcal{T}^{\alpha
\beta}-\mathcal{T}\mathcal{T}_{\xi\gamma}+2\mathcal{T}_{\xi}^{
\alpha}\mathcal{T}_{\gamma\alpha}.
\end{eqnarray}
The action of EMSG with scalar field becomes
\begin{equation}\label{3}
\mathcal{S}=\int
\sqrt{-\mathrm{g}}\left(\frac{f\left(\mathbb{R},\mathbf{T}^{2}
\right)}{2\kappa^2}+\mathcal{L}_{\mathrm{M}}+\mathcal{L}_{\varphi}
\right)d^4x,
\end{equation}
where
$\mathcal{L}_{\varphi}=\frac{\varepsilon}{2}\mathrm{g}^{\xi\gamma}
\partial_{\xi}\varphi\partial_{\gamma}\varphi-\vartheta(\varphi)$
is the scalar-Lagrangian and $\vartheta(\varphi)$ defines the
potential energy whereas $\varepsilon=1, -1$ represent the
\emph{quintessence} and \emph{phantom} models, respectively. For a
complete analysis, we investigate the effects of both positive and
the negative kinetic energy in the scalar field models. It is
expected that some important findings will be obtained to analyze
the mysterious universe due to the scalar field and EMSG.

We consider perfect matter distribution as
\begin{equation}\label{4}
\mathcal{T}_{\xi\gamma}=(\sigma+\varrho)\mathcal{V}_{\xi}\mathcal{V}_{
\gamma}+\varrho\mathrm{g}_{\xi\gamma},
\end{equation}
where four-velocity, energy density and  pressure of the fluid are
determined by $\mathcal{V}_{\xi}$, $\sigma$ and $\varrho$,
respectively. The homogeneous and isotropic spacetime is given as
\begin{equation}\label{5}
ds^{2}= -d\mathrm{t}^{2}+\mathrm{a}^{2}(\mathrm{t})(d\mathrm{x}^{2}
+d\mathrm{y}^{2}+d\mathrm{z}^{2}),
\end{equation}
where $\mathrm{a}(\mathrm{t})$ is the cosmic scale parameter. We use
the Lagrange multiplier approach to obtain point-like Lagrangian as
\begin{equation}\label{6}
\mathcal{S}= \int
(f-(\mathbb{R}-\bar{\mathbb{R}})\lambda_{1}-(\mathbf{T}^{2}-\mathbf
{\bar{T}}^{2})\lambda_{2}+\varrho+\frac{\varepsilon}{2}g^{\xi\gamma}
\partial_{\xi}\varphi\partial_{\gamma}\varphi-\vartheta)
\sqrt{-\mathrm{g}}d\mathrm{t},
\end{equation}
here
\begin{equation}\nonumber
\bar{\mathbb{R}}=
6\left(\frac{\ddot{\mathrm{a}}}{\mathrm{a}}+\frac{\dot{\mathrm{a}}^{2}
}{\mathrm{a}^{2}}\right), \quad \bar{\mathbf{T}}^{2}=
3\varrho^{2}+\sigma^{2}, \quad \lambda_{1}= f_{\mathbb{R}}, \quad
\lambda_{2}= f_{\mathbf{T}^{2}}.
\end{equation}
The corresponding Lagrangian becomes
\begin{eqnarray}\nonumber
\mathrm{L}&=&
\mathrm{a}^{3}(f-\mathbb{R}f_{\mathbb{R}}-\mathbf{T}^{2}f_{\mathbf{T}
^{2}}+\left(3\varrho^{2}+\sigma^{2}\right)f_{\mathbf{T}^{2}}-\vartheta
-\frac{\varepsilon\dot{\varphi}^{2}}{2}+\varrho)\\\label{7}&-&
6\mathrm{a}^{2}\dot{\mathrm{a}}(\dot{\mathrm{a}}\mathrm{a}^{-1}f_{
\mathbb{R}}+\dot{\mathbb{R}}f_{\mathbb{R}\mathbb{R}}+\dot{\mathbf{T}}^{2}
f_{\mathbb{R}\mathbf{T}^{2}}),
\end{eqnarray}
where dot is the derivative corresponding to cosmic time. The
Euler-Lagrange equations of motion, conjugate momenta $(P_{i})$ and
Hamiltonian $(\mathrm{H})$ are the key factors to explain the basic
characteristics of the system, determined as
\begin{equation}\label{8}
\frac{\partial
\mathrm{L}}{\partial\mathrm{q}^{i}}-\frac{dP_{i}}{d\mathrm{t}}=0,
\quad P_{i}=\left(\frac{\partial\mathrm{L}}{\partial\dot{\mathrm{q}}
^{i}}\right), \quad \mathrm{H}=\dot{q}^{i}P_{i}-\mathrm{L}, \quad
i=1,2,3,4,
\end{equation}
where
$\mathrm{q}_{i}=\mathrm{q}(\mathrm{a},\varphi,\mathbb{R},\mathbf{T}^2)$
are the generalized coordinates. Using Lagrangian (\ref{7}), the
conjugate momenta turn out to be
\begin{eqnarray}\nonumber
P_{\mathrm{a}}&=&-12\mathrm{a}\dot{\mathrm{a}}f_{\mathbb{R}}-6\mathrm{a}^{2}
\dot{\mathbb{R}}f_{\mathbb{R}\mathbb{R}}-6\mathrm{a}^{2}\dot{\mathbf{T}}^2
f_{\mathbb{R}\mathbf{T}^{2}}, \quad
P_{\mathbb{R}}=-6\mathrm{a}^{2}\dot{\mathbb{R}}f_{\mathbb{R}\mathbb{R}},
\\\nonumber
P_{\mathbf{T}^{2}}&=&-6\mathrm{a}^{2}\dot{\mathbf{T}}^{2}f_{\mathbb{R}
\mathbf{T}^{2}}, \quad
P_{\varphi}=-\mathrm{a}^{3}\varepsilon\dot{\varphi}.
\end{eqnarray}
The dynamical equations of the system become
\begin{eqnarray}\nonumber
&&\frac{1}{2}\left(f-\mathbb{R} f_{\mathbb{R}}-\mathbf{T}^{2}
f_{\mathbf{T}^{2}}+\left(3\varrho^{2}+\sigma^{2}\right)f_{
\mathbf{T}^{2}}+\varrho-\frac{\varepsilon\dot{\varphi}^{2}}{2}
\right)+2\frac{\dot{\mathrm{a}}}{\mathrm{a}}
\dot{f_{\mathbb{R}}}\\\label{9}&&-\frac{\vartheta}{2}+\frac{\mathrm{a}}{6}\Big\{
\left(6\varrho\varrho_{,\mathrm{a}}+2\sigma\sigma_{,\mathrm{a}}
\right)f_{\mathbf{T}^{2}}+\varrho_{,\mathrm{a}}\Big\}+\left(
\frac{\dot{\mathrm{a}}^{2}}{\mathrm{a}^{2}}+2\frac{\ddot{\mathrm
{a}}}{\mathrm{a}}\right)f_{\mathbb{R}}+\ddot{f_{\mathbb{R}}}=0,
\\\label{10}&&\left(\mathbb{R}-6\frac{\dot{\mathrm{a}}^{2}}{
\mathrm{a}^{2}}-6\frac{\ddot{\mathrm{a}}}{\mathrm{a}}\right)
f_{\mathbb{R}\mathbb{R}}+\left(\mathbf{T}^{2}-3\varrho^2-\sigma
^{2}\right)f_{\mathbb{R}\mathbf{T}^{2}}=0,\\\label{11}&&\left
(\mathbb{R}-6\frac{\dot{\mathrm{a}}^{2}}{\mathrm{a}^{2}}-6\frac
{\ddot{\mathrm{a}}}{\mathrm{a}}\right)f_{\mathbb{R}\mathbf{T}^{2}}
+\left(\mathbf{T}^{2}-3\varrho^2-\sigma^{2}\right)f_{\mathbf{T}^{2}
\mathbf{T}^{2}}=0,
\\\label{12}&& \mathrm{a}\ddot{\varphi}+3\varepsilon\dot{\mathrm{a}}
\dot{\varphi}-\mathrm{a}\vartheta'(\varphi)=0.
\end{eqnarray}
We establish the Hamiltonian to evaluate the total systematic energy
as
\begin{eqnarray}\nonumber
\mathrm{H}&=&-\mathrm{a}^{3}\left(f-\mathbb{R}f_{\mathbb{R}}-\mathbf{T}^{2}
f_{\mathbf{T}^{2}}+\left(3\varrho^{2}+\sigma^{2}\right)f_{\mathbf{T}^{2}}
+\varrho-\frac{\varepsilon\dot{\varphi}^{2}}{2}-\vartheta\right)
\\\label{13}&-&6\mathrm{a}^{2}\dot{\mathrm{a}}\left(\dot{\mathbb{R}}f_{
\mathbb{R}\mathbb{R}}+\dot{\mathbf{T}}^{2}f_{\mathbb{R}\mathbf{T}^{2}}
\right)-6\mathrm{a}\dot{\mathrm{a}}^{2}f_{\mathbb{R}}.
\end{eqnarray}
The dynamical equations (\ref{9})-(\ref{12}) are very complicated
because of multivariate function and their derivatives. These
equations are extremely difficult to solve directly but can be
solved in one of the two ways. The first is to solve them using an
appropriate analytic or numeric method, while the second is to use
the NS methodology to find exact solutions. Although this is a
non-conserved theory whereas NS technique allows us to obtain
conserved parameters, which are then used to investigate the
enigmatic cosmos. As a result, the later method is more intriguing,
and we will use it here.

\section{Noether Symmetries}

The NS approach provides an intriguing method to construct new
cosmological models and corresponding geometries in alternative
gravitational theories. This section examines the Noether equations
of isotropic spacetime with the minimal coupling of scalar field and
matter in EMSG. This strategy generates a vector field that
corresponds to tangent space. Consequently, the vector field serves
as a symmetry generator and yields first integrals that can be used
to examine physically viable solutions. The NS technique and
corresponding first integrals are strongly motivated from the Lie
symmetries, i.e., the NS for a given Lagrangian exists if the Lie
derivative $(\mathbb{L})$ of the Lagrangian corresponding to vector
field $\mathbb{K}$ is zero ($\mathbb{L}_{\mathbb{K}}\mathrm{L}=0)$.
The Noether generators are defined as
\begin{equation}\label{14}
\mathbb{K}= \tau(\mathrm{t},\mathrm{a}, \mathbb{R}, \mathbf{T}^{2},
\varphi)\frac{\partial}{\partial\mathrm{t}}+\eta^{i}(\mathrm{t},\mathrm{a},
\mathbb{R}, \mathbf{T}^{2},
\varphi)\frac{\partial}{\partial\mathrm{q}^{i}},
\end{equation}
where $\tau$ and $\eta^{i}=(\alpha, \beta, \chi, \delta)$ are the
unknown parameters of $\mathbb{K}$. The Lagrangian must fulfill the
invariance condition for the existence of NS, expressed as
\begin{equation}\label{15}
\mathbb{K}^{[1]}\mathrm{L}+(\mathbb{D}\tau)\mathrm{L}=
\mathbb{D}\psi,
\end{equation}
where $\psi$ is the boundary term,
$\mathbb{D}=\frac{\partial}{\partial
\mathrm{t}}+\dot{\mathrm{q}}^{i}\frac{\partial}{\partial
\mathrm{q}^{i}}$ defines the total derivative and
$\mathbb{K}^{[1]}=\mathbb{K}+\dot{\eta}^{i}\frac{\partial}{\partial
\dot{\mathrm{q}}^{i}}$ is the first-order prolongation. To analyze
the symmetry generators with conserved quantities, the first order
prolongation is given by
\begin{eqnarray}\nonumber
\mathbb{K}^{[1]}&=&\tau\frac{\partial}{\partial\mathrm{t}}+(\alpha+\dot{\alpha})\frac{\partial}
{\partial\mathrm{a}}+(\beta+\dot{\beta})\frac{\partial}{\partial\mathbb{R}}+(\chi+\dot{\chi})
\frac{\partial}{\partial\mathbf{T}^{2}}+(\delta+\dot{\delta})\frac{\partial}{\partial
\varphi},
\end{eqnarray}
where
\begin{eqnarray}\nonumber
\dot{\alpha}&=&\frac{\partial
\alpha}{\partial\mathrm{t}}+\dot{\mathrm{a}}\frac{\partial
\alpha}{\partial \mathrm{a}}+ \dot{\mathbb{R}}\frac{\partial
\alpha}{\partial \mathbb{R}}+ \dot{\mathbf{T}}^{2}\frac{\partial
\alpha}{\partial\mathbf{T}^{2}}-\dot{\mathrm{a}}\Omega,
\\\nonumber
\dot{\beta}&=&\frac{\partial
\beta}{\partial\mathrm{t}}+\dot{\mathrm{a}}\frac{\partial
\beta}{\partial \mathrm{a}}+ \dot{\mathbb{R}}\frac{\partial
\beta}{\partial \mathbb{R}}+ \dot{\mathbf{T}}^{2}\frac{\partial
\beta}{\partial \mathbf{T}^{2}}-\dot{\mathbb{R}}\Omega,
\\\nonumber
\dot{\chi}&=&\frac{\partial \chi}{\partial
\mathrm{t}}+\dot{\mathrm{a}}\frac{\partial \chi}{\partial
\mathrm{a}}+ \dot{\mathbb{R}}\frac{\partial \chi}{\partial
\mathbb{R}}+ \dot{\mathbf{T}}^{2}\frac{\partial \chi}{\partial
\mathbf{T}^{2}}-\dot{\mathbf{T}}^{2}\Omega,
\\\nonumber
\dot{\delta}&=&\frac{\partial \delta}{\partial
\mathrm{t}}+\dot{\mathrm{a}}\frac{\partial \delta}{\partial
\mathrm{a}}+ \dot{\mathbb{R}}\frac{\partial \delta}{\partial
\mathbb{R}}+ \dot{\mathbf{T}}^{2}\frac{\partial \delta}{\partial
\mathbf{T}^{2}}-\dot{\varphi}\Omega,
\end{eqnarray}
here $\Omega=\left(\frac{\partial \tau}{\partial
\mathrm{t}}+\dot{\mathrm{a}}\frac{\partial \tau}{\partial
\mathrm{a}}+ \dot{\mathbb{R}}\frac{\partial \tau}{\partial
\mathbb{R}}+ \dot{\mathbf{T}}^{2}\frac{\partial \tau}{\partial
\mathbf{T}^{2}}\right)$.

The corresponding conserved parameters are expressed as
\begin{equation}\label{17}
\mathcal{I}= \eta^{i}\frac{\partial
\mathrm{L}}{\partial\dot{q}^{i}}-\tau \mathcal{H}-\psi.
\end{equation}
This is the significant part of NS that plays a crucial role to
derive viable solutions and is also known as the first integral of
motion. By comparing the coefficients of Eq.(\ref{15}), we obtain
\begin{eqnarray}\label{19}
&&6\mathrm{a}^{2}\alpha_{,\mathrm{t}}f_{\mathbb{R}\mathbb{R}}
+\psi_{,\mathbb{R}}=0, \quad
6\mathrm{a}^{2}\alpha_{,\varphi}f_{\mathbb{R}\mathbb{R}}+
\mathrm{a}^{3}\varepsilon\delta_{,\mathbb{R}}=0,
\\\label{20}
&&12\mathrm{a}\alpha_{,\mathbf{T}^{2}}f_{\mathbb{R}}+6\mathrm{a}
^{2}\beta_{,\mathbf{T}^{2}}f_{\mathbb{R}\mathbb{R}}+6\mathrm{a}^{2}
\chi_{,\mathbf{T}^{2}}f_{\mathbb{R}\mathbf{T}^{2}}+\psi_{,\mathrm{a}}
=0,\\\label{21}
&&\mathrm{a}^{3}\varepsilon\delta_{,\mathrm{a}}+6\mathrm{a}^{2}
\chi_{,\varphi}f_{\mathbb{R}\mathbf{T}^{2}}+6\mathrm{a}^{2}
\beta_{,\varphi}f_{\mathbb{R}\mathbb{R}}+12\mathrm{a}\alpha_{,\varphi}
f_{\mathbb{R}}=0,\\\label{22}&&\alpha-\tau_{,\mathrm{t}}-2\mathrm{a}^{3}
\delta_{,\varphi}+2\mathrm{a}^{3}\tau_{,\mathrm{t}}=0, \quad
6\mathrm{a}^{2}\alpha_{,\mathrm{t}}f_{\mathbb{R}\mathbf{T}^{2}}
+\psi_{,\mathbf{T}^{2}}=0,
\\\label{23}&&\mathrm{a}\tau_{,\mathrm{a}}f_{\mathbb{R}}=0,\quad
\mathrm{a}\tau_{,\mathbb{R}}f_{\mathbb{R}} =0,\quad
\mathrm{a}\tau_{,\mathbf{T}^{2}}f_{\mathbb{R}} =0, \quad
\mathrm{a}\tau_{,\varphi}f_{\mathbb{R}} =0,
\\\label{24}
&&6\mathrm{a}^{2}\tau_{,\mathbb{R}}f_{\mathbb{R}\mathbb{R}} =0,\quad
6\mathrm{a}^{2}\tau_{,\mathbf{T}^{2}}f_{\mathbb{R}\mathbf{T}^{2}}
=0,\quad
\mathrm{a}^{3}\varepsilon\delta_{,\mathbf{T}^{2}}+\psi_{,\varphi}=0,
\\\label{25}
&&6\mathrm{a}^{2}\alpha_{,\varphi}f_{\mathbb{R}\mathbf{T}^{2}}+
\mathrm{a}^{3}\varepsilon\delta_{\mathbf{T}^{2}}=0, \quad
6\mathrm{a}^{2}\alpha_{,\mathbf{T}^{2}}f_{\mathbb{R}\mathbb{R}}
+6\mathrm{a}^{2}\alpha_{,\mathbb{R}}f_{\mathbb{R}\mathbf{T}^{2}} =0,
\\\nonumber
&&\alpha f_{\mathbb{R}}+\alpha \beta
f_{\mathbb{R}\mathbb{R}}+\mathrm{a}\chi
f_{\mathbb{R}\mathbf{T}^{2}}-\mathrm{a}\tau_{,\mathrm{t}}+2\mathrm{a}
\alpha_{,\mathrm{a}}f_{\mathbb{R}}+\mathrm{a}^{2}\beta_{,\mathrm{a}}
f_{\mathbb{R}\mathbb{R}}
\\\label{26}
&&+\mathrm{a}^{2}\chi_{,\mathrm{a}}f_{\mathbb{R}\mathbf{T}^{2}}=0,
\\\nonumber
&&2\mathrm{a}\alpha_{,\mathbb{R}}f_{\mathbb{R}}+2\alpha
f_{\mathbb{R}\mathbb{R}}+\mathrm{a}\alpha_{,\mathbb{R}}
f_{\mathbb{R}\mathbb{R}}+\alpha \beta
f_{\mathbb{R}\mathbb{R}\mathbb{R}}+\mathrm{a}\chi
f_{\mathbb{R}\mathbb{R}\mathbf{T}^{2}}-\mathrm{a}
\tau_{,\mathrm{t}}f_{\mathbb{R}\mathbb{R}}
\\\label{27}
&&+\mathrm{a}\chi_{,\mathbb{R}}f_{\mathbb{R}\mathbf{T}^{2}}
+\mathrm{a}\beta_{,\mathbb{R}}f_{\mathbb{R}\mathbb{R}}=0,
\\\nonumber
&&(2\alpha+\mathrm{a}\alpha_{,
\mathrm{a}}+\mathrm{a}\chi_{,\mathbf{T}^{2}}-\mathrm{a}
\tau_{,\mathrm{t}}) f_{\mathbb{R}\mathbf{T}^{2}}+\alpha \beta
f_{\mathbb{R}\mathbb{R}\mathbf{T}^{2}}+\mathrm{a}\chi
f_{\mathbb{R}\mathbf{T}^{2}\mathbf{T}^{2}}+2\mathrm{a}
\alpha_{,T}f_{\mathbb{R}}
\\\label{28}
&&+\mathrm{a}\beta_{,\mathbf{T}^{2}}f_{\mathbb{R}\mathbb{R}}=0,
\\\nonumber
&&3\mathrm{a}^{2}\alpha(f-\mathbb{R}f_{\mathbb{R}}-\mathbf{T}^{2}
f_{\mathbf{T}^{2}}+(3\varrho^{2}+\sigma^{2})f_{\mathbf{T}^{2}}
-\vartheta+\varrho)+\mathrm{a}^{3}\alpha\varrho_{a}+2\mathrm{a}^{3}\alpha
\\\nonumber&&
\times(3\varrho\varrho_{\mathrm{a}}+
\sigma_{\mathrm{a}})f_{\mathbf{T}^{2}}+\mathrm{a}^{3}\beta
(-\mathbb{R}f_{\mathbb{R}\mathbb{R}}-\mathbf{T}^{2}
f_{\mathbb{R}\mathbf{T}^{2}})+\mathrm{a}^{3}\beta
f_{\mathbb{R}\mathbf{T}^{2}})(3\varrho^{2}+\sigma^{2})
\\\nonumber&&
+\mathrm{a}^{3}\chi(-\mathbb{R}
f_{\mathbb{R}\mathbf{T}^{2}}-\mathbf{T}^{2}f_{\mathbf{T}^{2}\mathbf{T}^{2}}
+(3\varrho^{2}+\sigma^{2})f_{\mathbf{T}^{2}\mathbf{T}^{2}})
+\mathrm{a}^{3}\vartheta'\delta
\\\label{29}&&
+\mathrm{a}^{3}\tau_{,\mathrm{t}}(f-\mathbb{R}
f_{\mathbb{R}}-\mathbf{T}^{2} f_{\mathbf{T}^{2}}
+(3\varrho^{2}+\sigma^{2})f_{\mathbf{T}^{2}}+\varrho)
-\psi_{,\mathrm{t}}=0.
\end{eqnarray}
These equations are the main ingredients to examine the dark
universe in the framework of EMSG. In the next section, we
manipulate the above system to obtain exact cosmological solutions
for various EMSG models.

\section{Exact Cosmological Solutions}

Here, we establish the symmetry generators with conserved parameters
and corresponding physical solutions. It is difficult to get exact
solutions of the above system of PDEs without using a particular
EMSG model due to its complexity and highly nonlinearity. We assume
various models that reduce the complexity of the system and help to
obtain cosmological solutions.

\subsection*{Case I: Minimal Coupling Model}

This case investigates EMSG by taking a minimal interaction between
curvature and matter as
$f(\mathbb{R},\mathbf{T}^{2})=\mathbb{R}^{m}+(\mathbf{T}^{2})^n$,
where $m$ and $n$ are arbitrary constants \cite{8}. Solving
Eqs.(\ref{19})-(\ref{29}), we get
\begin{eqnarray}\nonumber
&&\alpha=0, \quad \beta=
\frac{\mathbb{R}^{2-m}}{\mathrm{a}}\left(\frac{c_{2}\varepsilon
\varphi}{6m(m-1)}+c_{3}+c_{4}\mathrm{a}\right), \quad \delta=
\frac{c_{2}\mathbb{R}}{\mathrm{a}}, \quad \tau=c_{1},
\\\label{30}&&
\vartheta=\frac{m\varphi(1-m)(c_{3}+\mathrm{a}c_{4})}{c_{2}}
-\frac{\varphi^{2}\varepsilon}{12}+c_{5}, \quad \psi=0.
\end{eqnarray}
where $c_{i}$ are arbitrary constants, whereas Eq.(\ref{20}) yields
\begin{eqnarray}\label{31}
\varphi= \mathrm{a}c_{6}-\frac{6c_{3}m(m-1)}{c_{2}\varepsilon}.
\end{eqnarray}
The boundary terms for this value of $\varphi$ reduce to
\begin{eqnarray}\nonumber
\vartheta&=& \frac{m(1-m)(c_{3}+\mathrm{a}c_{4})}{c_{2}}
\left(\mathrm{a}c_{6}-\frac{6c_{3}m(m-1)}{c_{2}\varepsilon}
\right)-\frac{\left(\mathrm{a}c_{6}-\frac{6c_{3}m(m-1)}{c_{2}
\varepsilon}\right)^{2}\varepsilon}{12}+c_{5},
\\\nonumber
\beta&=&\frac{\mathbb{R}^{2-m}}{\mathrm{a}}\left(\frac{\mathrm{a}c_{2}c_{6}
\varepsilon}{6m(m-1)}+c_{4}\mathrm{a}\right).
\end{eqnarray}
The symmetry generators become
\begin{eqnarray}\nonumber
\mathbb{K}_{1}=\frac{\partial}{\partial \mathrm{t}}, \quad
\mathbb{K}_{2}=\frac{\mathbb{R}^{2-m}\varepsilon\varphi}
{6\mathrm{a}m(m-1)}\frac{\partial}{\partial\mathbb{R}}+
\frac{c_{2}\mathbb{R}}{\mathrm{a}}\frac{\partial}{\partial \varphi},
\quad \mathbb{K}_{3}=\frac{\mathbb{R}^{2-m}}
{\mathrm{a}}\frac{\partial}{\partial\mathbb{R}},\quad
\mathbb{K}_{4}= \mathbb{R}^{2-m}\frac{\partial}{\partial
\mathbb{R}},
\end{eqnarray}
and the corresponding conserved quantities are
\begin{eqnarray}\nonumber
\mathcal{I}_{1}&=&\mathrm{a}^{3}\mathbb{R}^{m}(1-m)+\mathrm{a}^{3}
(\mathbf{T}^{2})^{n}-0.5\mathrm{a}^{3}\varepsilon\dot{\varphi}^{2}
+0.08\mathrm{a}^{3}\varphi^{2}\varepsilon-\mathrm{a}^{3}c_{5}
\\\nonumber
&+&6\mathrm{a}^{2}\dot{\mathrm{a}}
\dot{\mathbb{R}}{\mathbb{R}}^{m-2}{m}(m-1)+6\mathrm{a}\dot{\mathrm{a}}
\mathbb{R}^{m-1}m, \quad
\mathcal{I}_{2}=-\left(\mathrm{a}\mathbb{R}+\dot{\mathrm{a}}\right)
\mathrm{a}\varepsilon\varphi,
\\\nonumber
\mathcal{I}_{3}&=&{\frac{\mathrm{a}m(c_{1}\mathrm{a}^{2}\varphi
m-c_{1}\mathrm{a}^{2} \varphi-6c_{2}
\dot{\mathrm{a}}m+6c_{2}\dot{\mathrm{a}})}{c_{2}}},
\\\nonumber
\mathcal{I}_{4}&=&\frac{\mathrm{a}^{2}m(c_{1}\mathrm{a}^{2}\varphi
m-c_{1}\mathrm{a}^{2}\varphi-6c_{2}\dot{\mathrm{a}}m+6c_{2}
\dot{\mathrm{a}})}{c_{2}}.
\end{eqnarray}
We substitute Eq.(\ref{31}) into (\ref{12}) to obtain the exact
solution
\begin{eqnarray}\label{34}
\mathrm{a}(\mathrm{t})&=&{6}^{\frac{1}{2(3\varepsilon+1)}}
\left[\frac{c_{2}c_{6}\varepsilon+6c_{4}m(m-1)}{c_{2}c_{6}
(3\varepsilon+1)\left(a_{1}\sin\mathbb{A}-a_{2}\cos\mathbb{A}
\right)^{2}}\right]^{\frac{-1}{2(3\varepsilon+1)}},
\end{eqnarray}
where $a_{1}$ and $a_{2}$ are integration constants and
\begin{eqnarray}\nonumber
\mathbb{A}={\frac{\sqrt{3\varepsilon+1}\sqrt{c_{2}c_{6}
\varepsilon+6c_{4}m(m-1)}\mathrm{t}}{\sqrt{6c_{2}c_{6}}}}.
\end{eqnarray}

To examine this solution, we analyze the graphical behavior of
\emph{scale factor}, \emph{Hubble}, \emph{deceleration} and $r-s$
\emph{parameters} that play a key role in the study of cosmology.
The Hubble parameter $(\mathcal{H}=\dot{\mathrm{a}}\mathrm{a}^{-1})$
explains the expansion rate of the universe that how fast universe
is expanding. This can be used to determine the age of the universe
and its history. The deceleration parameter
$(\mathrm{q}=-\mathcal{H}\mathcal{H}^{-2}-1)$ determines the
behavior of cosmic expansion, either it is constant
$(\mathrm{q}=0)$, decelerated $(\mathrm{q}>0)$ or accelerated
$(\mathrm{q}<0)$ expansion phase. The Hubble and deceleration
parameters for homogeneous and isotropic spacetime become
\begin{eqnarray}\label{35}
\mathcal{H}&=&-{\frac{\sqrt{c_{2}c_{6}\varepsilon+6c_{4}m(m-1)}}
{\sqrt{6c_{2}c_{6}(3\varepsilon+1)}}}\left[\frac{a_{1}\cos\mathbb{A}+
a_{2}\sin\mathbb{A}}{\mathrm{a}_{{2}}\cos\mathbb{A}-\mathrm{a}_{{1}}
\sin\mathbb{A}}\right],
\\\label{36}
\mathrm{q}&=&\frac{a_{1}^{2}+3\varepsilon(a_{1}^{2}+a_{2}^{2})
-(a_{1}^{2}+a_{2}^{2})\cos^{2}\mathbb{A}-2a_{1}a_{2}\sin\mathbb{A}
\cos\mathbb{A}}{a_{1}^{2}\cos^{2}\mathbb{A}-a_{2}^{2}\cos^{2}
\mathbb{A}-2a_{1}a_{2}\sin\mathbb{A}\cos\mathbb{A}+a_{2}^{2}}.
\end{eqnarray}
The $r-s$ parameters investigate the attributes of DE by
constructing a relation among formulated and standard cosmic models,
defined as
\begin{eqnarray}\nonumber
r=\mathrm{q}+2\mathrm{q}^{2}-\frac{\dot{\mathrm{q}}}{\mathcal{H}},
\quad s=\frac{r-1}{3(\mathrm{q}-\frac{1}{2})}.
\end{eqnarray}
The constructed model corresponds to $\Lambda$CDM model for $(r, s)
= (1, 0)$, whereas $s>0$ and $r<1$ represent the quintessence and
phantom eras of DE, respectively. In this work, these quantities are
given in appendix \textbf{A}.

The graphical representation of scale factor and Hubble parameter
for the proposed model corresponding to quintessence and phantom
models is given in Figures \textbf{1} and \textbf{2}, respectively.
Figure \textbf{1} exhibits that the scale factor is positive and
monotonically increasing corresponding to the quintessence model
which determines the current cosmic expansion while the phantom
model explains the decelerated universe. The Hubble parameter is
positive corresponding to both models which represents the cosmic
expansion as given in Figure \textbf{2}. Figure \textbf{3} indicates
that deceleration parameter is negative for $\varepsilon=1$ and
$\varepsilon=-1$ which describes that our cosmos is in the expansion
phase. The scalar field shows cosmic expansion for the quintessence
model while the phantom model represents expanding phase of the
universe for a small value of the model parameter. Figure \textbf{5}
(left plot) shows that $r-s$ parameters support quintessence and
phantom phases of DE whereas the right plot describes accelerated
expansion of the universe. In this case, the solutions obtained for
$\varepsilon=1$ are consistent with current observations implying
that EMSG supports current cosmic expansion.
\begin{figure}
\epsfig{file=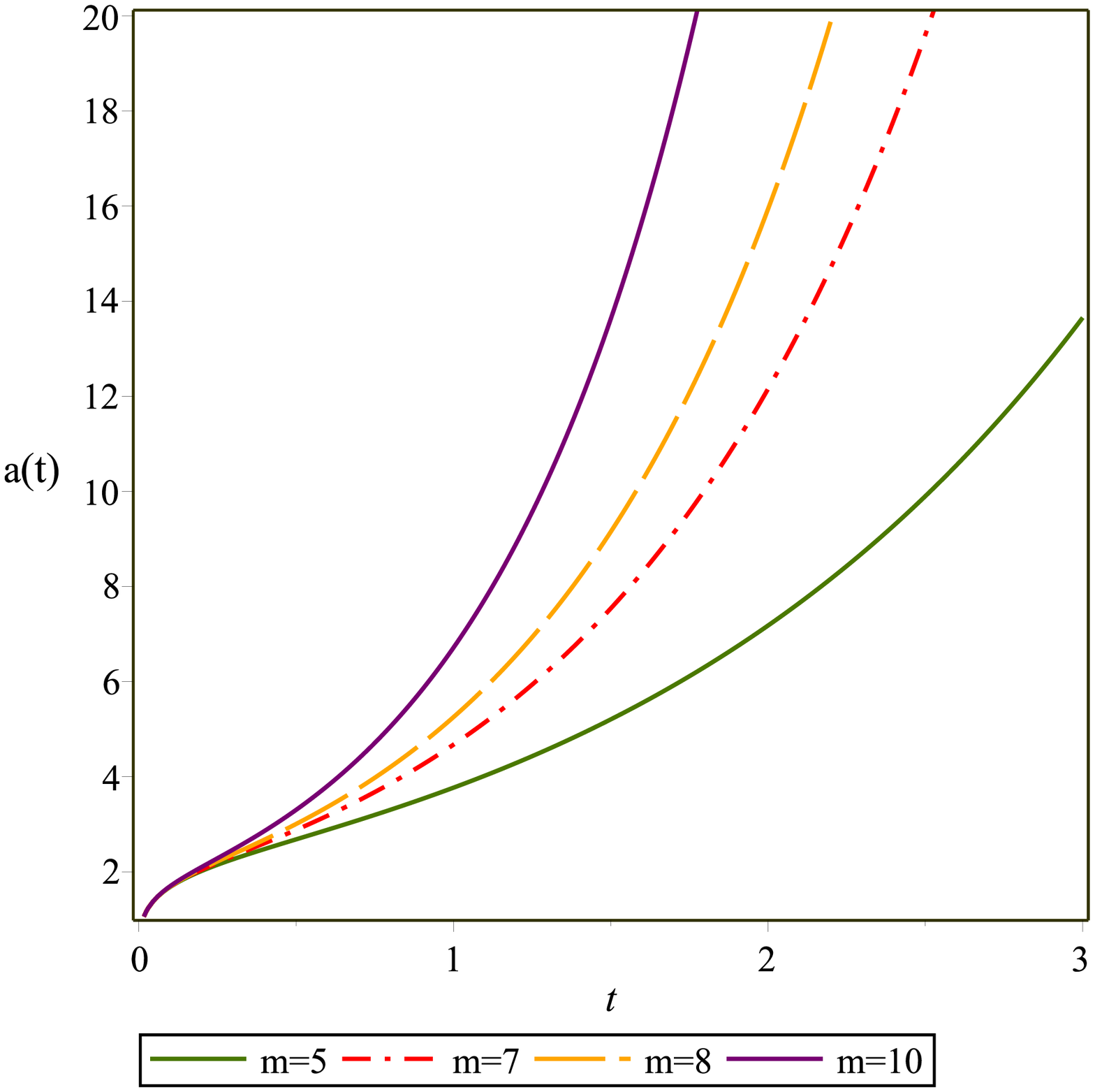,width=.5\linewidth}
\epsfig{file=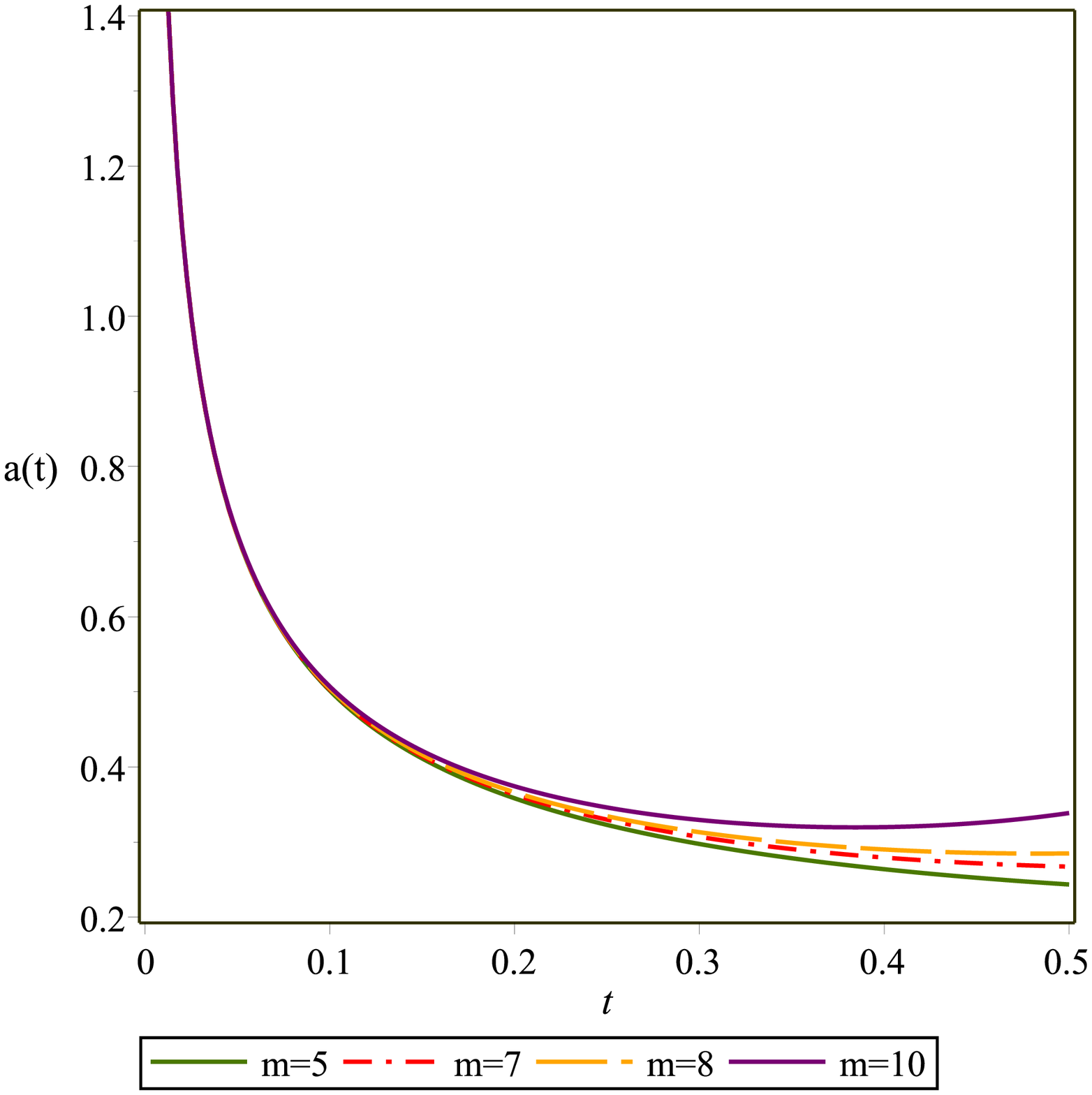,width=.5\linewidth} \caption{Behavior of scale
parameters for $\varepsilon=1$ (Left) and $\varepsilon=-1$ (Right).}
\end{figure}
\begin{figure}
\epsfig{file=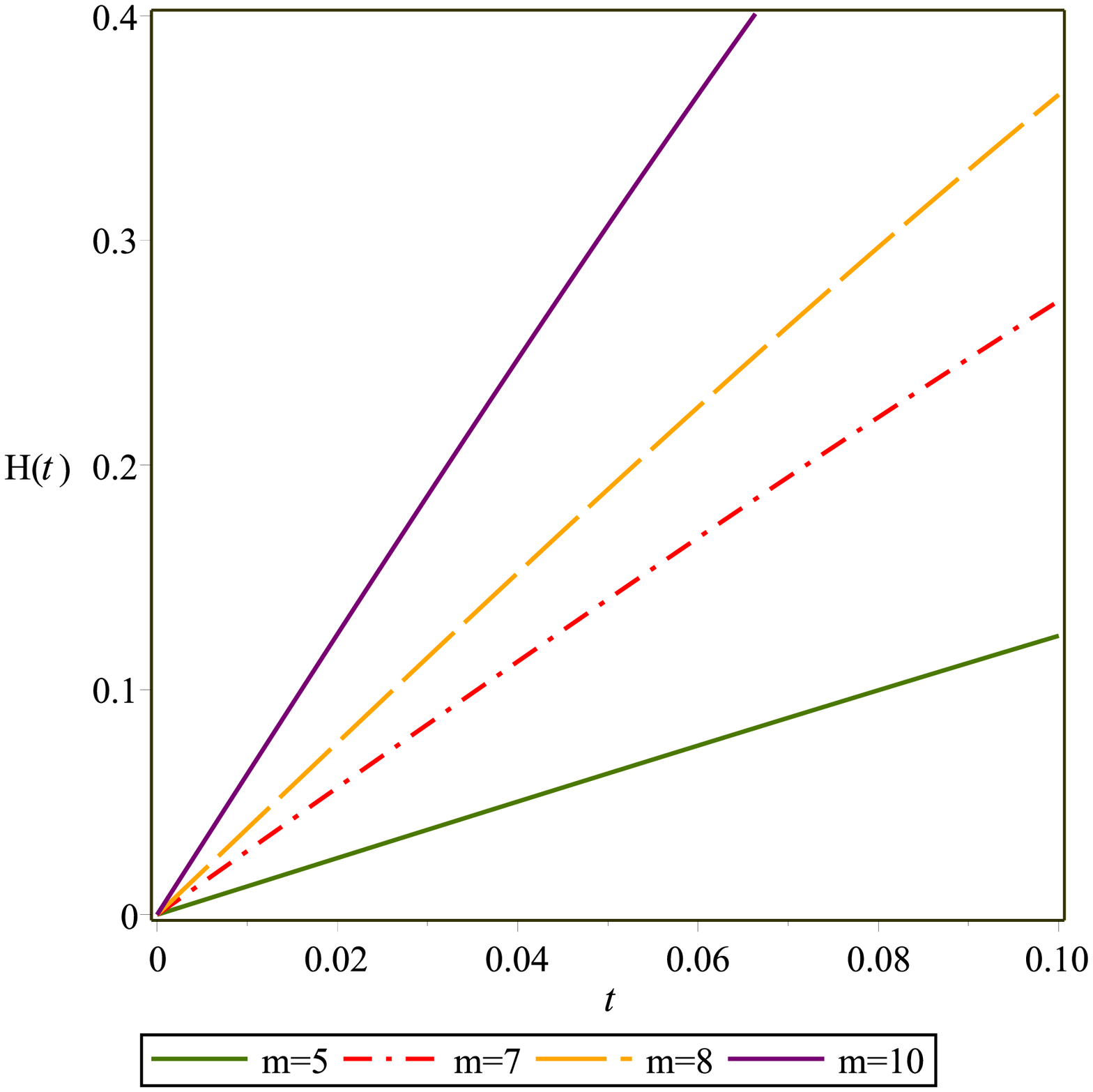,width=.5\linewidth}
\epsfig{file=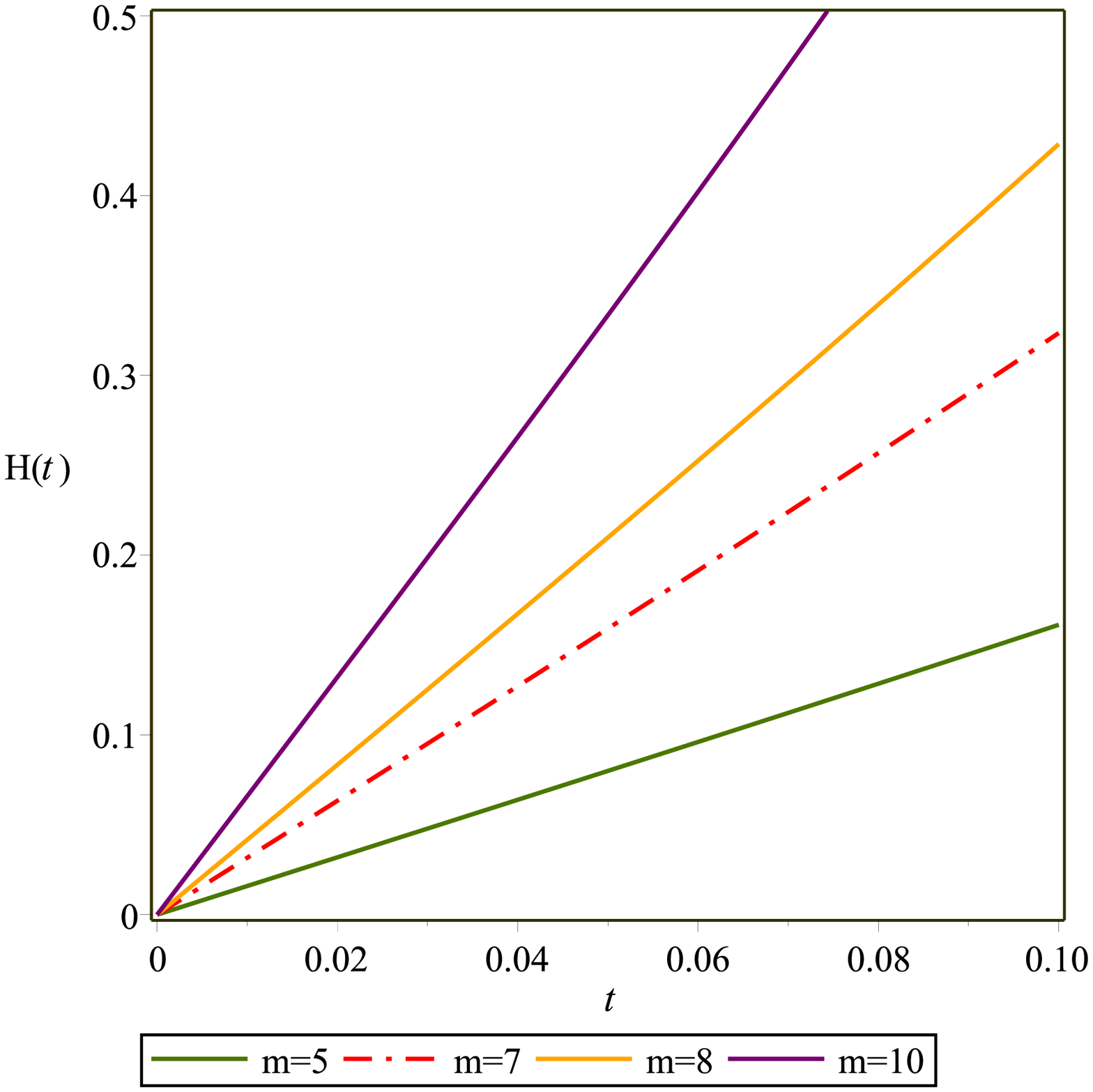,width=.5\linewidth} \caption{Plots of Hubble
Parameter for $\varepsilon=1$ (Left) and $\varepsilon=-1$ (Right).}
\end{figure}
\begin{figure}
\epsfig{file=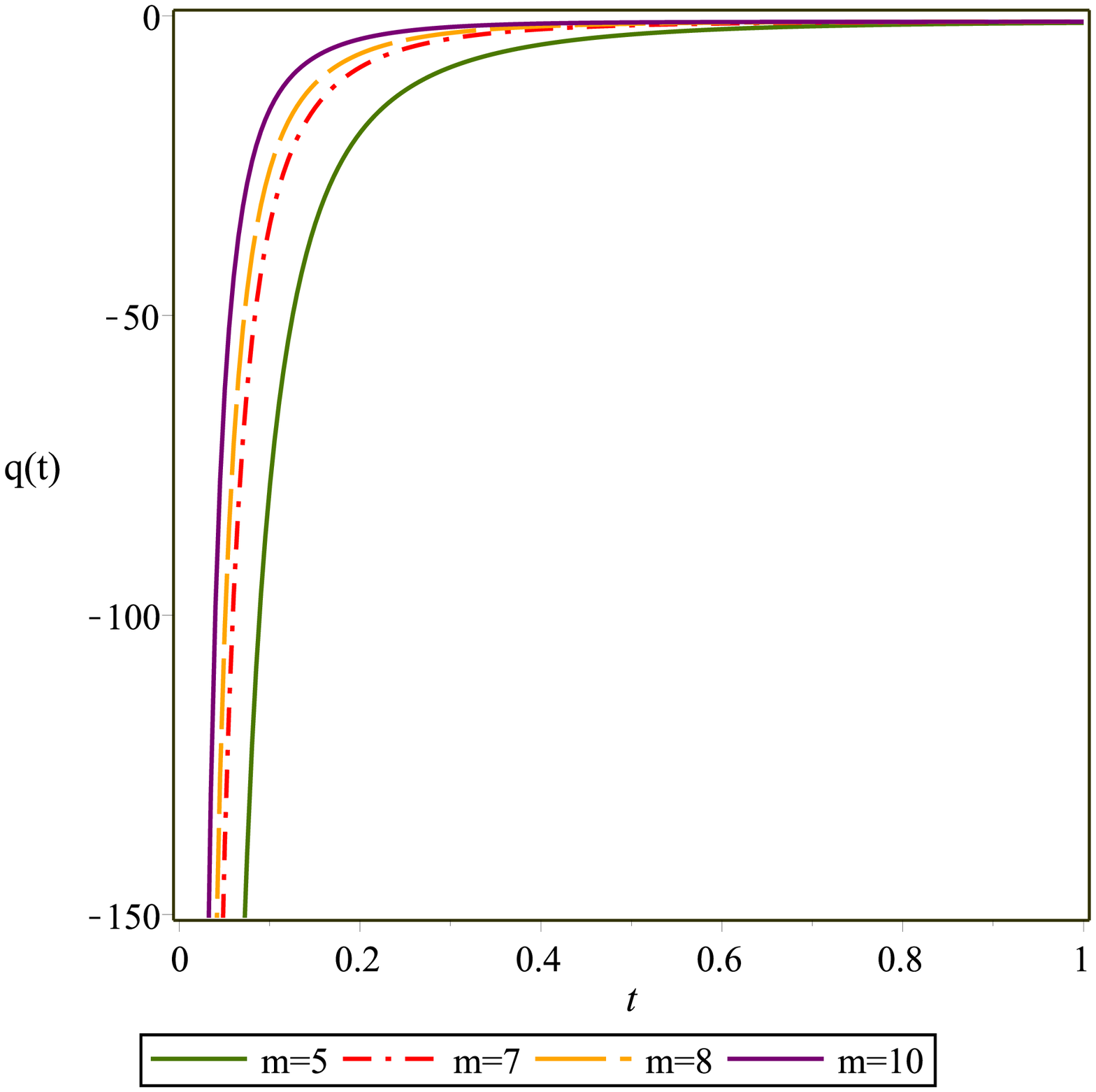,width=.5\linewidth}
\epsfig{file=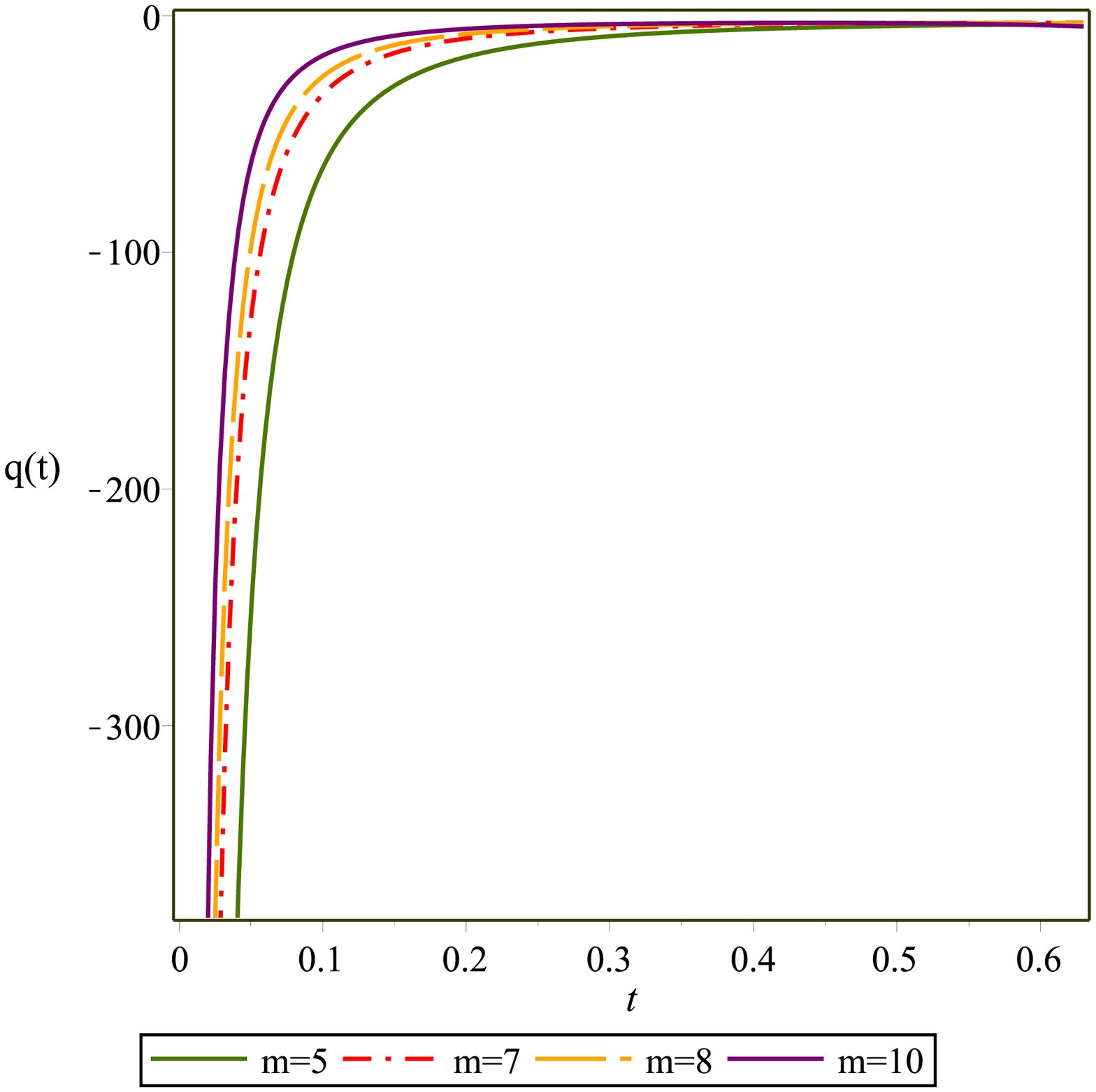,width=.5\linewidth} \caption{Deceleration
Parameter for $\varepsilon=1$ (Left) and $\varepsilon=-1$ (Right).}
\end{figure}
\begin{figure}
\epsfig{file=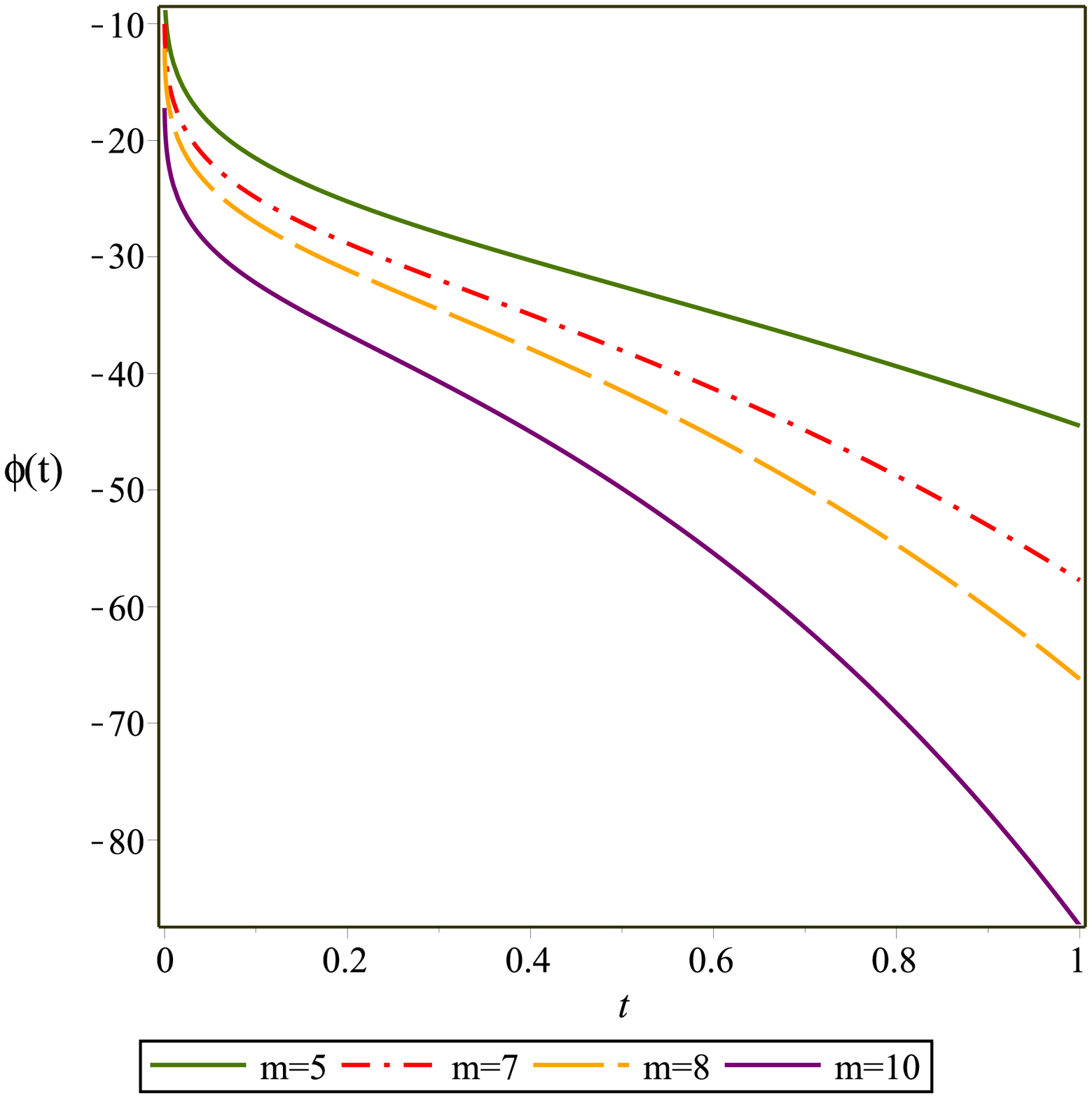,width=.5\linewidth}
\epsfig{file=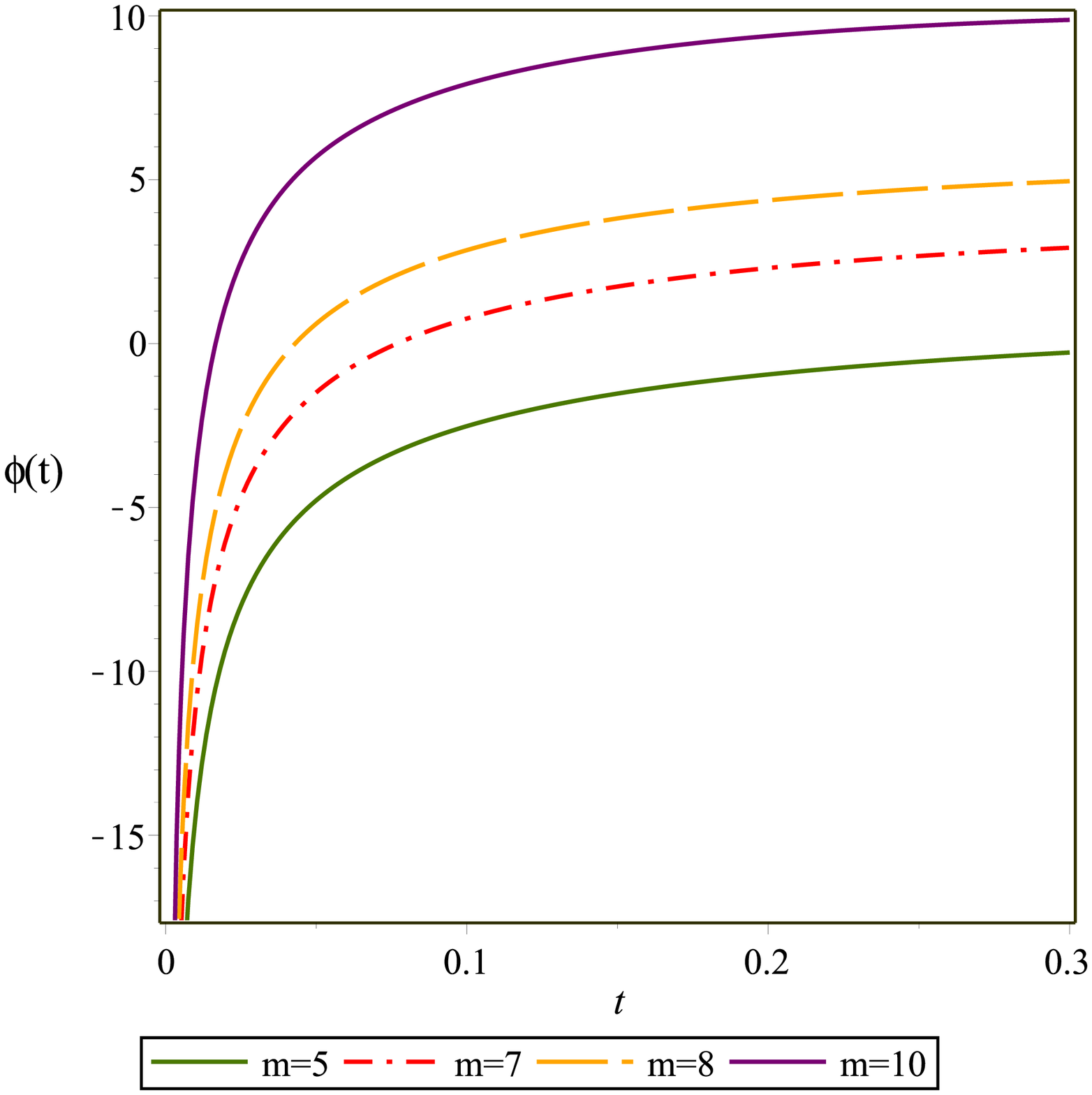,width=.5\linewidth} \caption{Graphs of scalar
field for $\varepsilon=1$ (Left) and $\varepsilon=-1$ (Right).}
\end{figure}
\begin{figure}
\epsfig{file=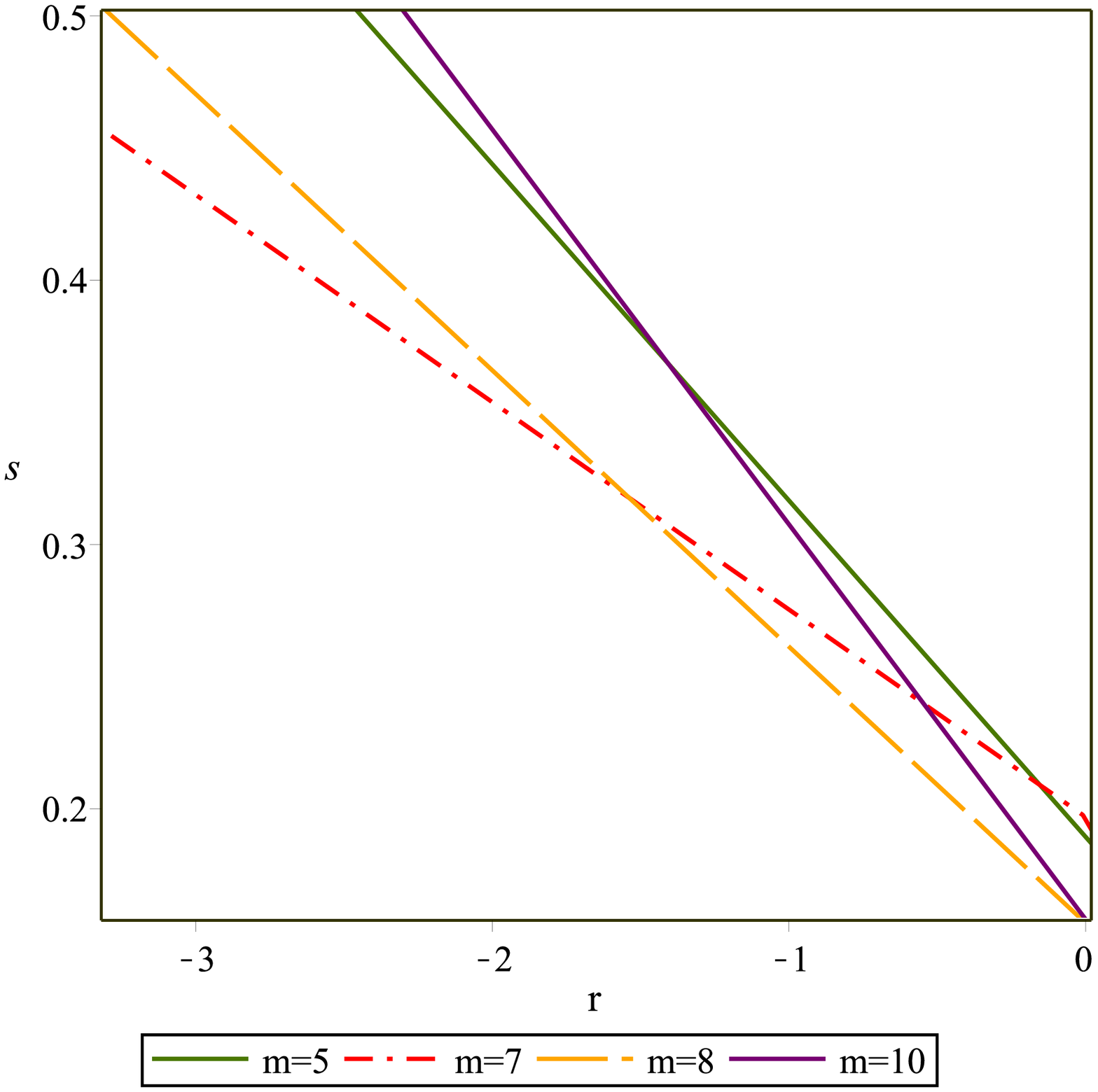,width=.5\linewidth}
\epsfig{file=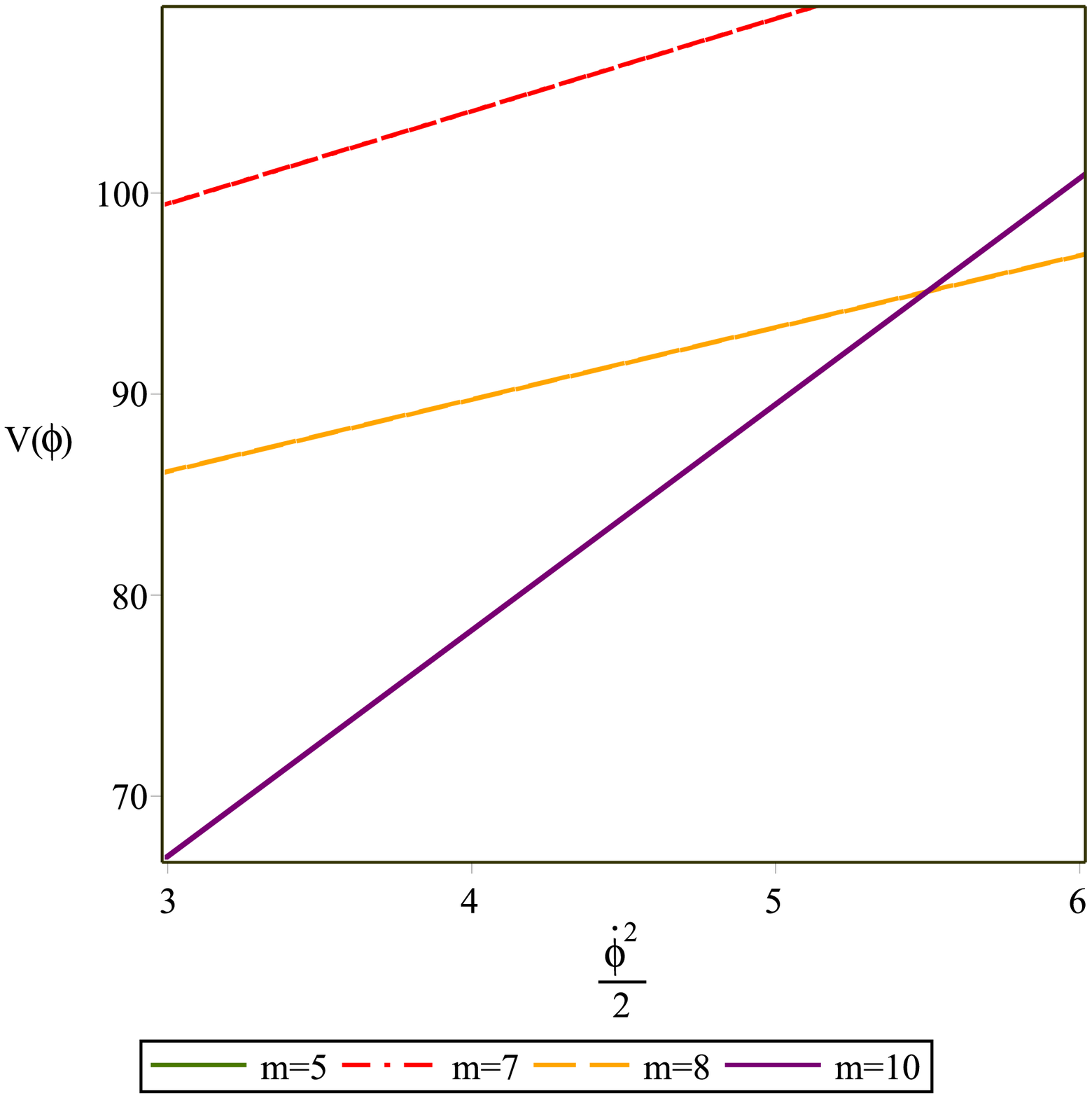,width=.5\linewidth} \caption{Plots of $r-s$
parameter (left) and potential energy (right) for $\varepsilon=1$}
\end{figure}

\subsection*{Case II: Non-Minimal Coupling Model}

This case examines the EMSG by considering
$f(\mathbb{R},\mathbf{T}^{2})=f_{0}\mathbb{R}\mathbf{T}^{2}$, where
$f_{0}$ is a constant \cite{8}. Solving Eqs.(\ref{19})-(\ref{29}),
we have
\begin{equation}\label{39}
\chi=\frac{c_{2}\varepsilon
\varphi}{6\mathrm{a}f_{0}}+\frac{c_{3}}{\mathrm{a}}, \quad
\delta=\frac{c_{2}\mathbb{R}}{\mathrm{a}}, \quad \tau=c_{1}, \quad
\vartheta=\frac{\varphi
c_{3}f_{0}}{c_{2}}-\frac{\varphi^{2}\varepsilon}{12}+c_{4},
\psi=0=\alpha.
\end{equation}
The symmetry generators with conserved quantities are
\begin{eqnarray}\label{40}
\mathbb{K}_{1}&=&\frac{\partial}{\partial \mathrm{t}}, \quad
\mathbb{K}_{2}=\frac{\varepsilon\varphi}{6\mathrm{a}f_{0}}
\frac{\partial}{\partial\mathbf{T}^{2}}+\frac{\mathbb{R}}
{\mathrm{a}}\frac{\partial}{\partial\varphi}, \quad \mathbb{K}_{3}=
\frac{1}{\mathrm{a}}\frac{\partial}{\partial \mathbf{T}^{2}},
\\\nonumber
\mathcal{I}_{1}&=&0.08\mathrm{a}^{3}\varphi^{2}\varepsilon
-0.5\mathrm{a}^{3}\varepsilon\dot{\varphi}^{2}-\mathrm{a}^{3}
c_{4}+6\mathrm{a}^{2}\dot{\mathrm{a}}\dot{\mathbf{T}}^{2}f_{0}
+6\mathrm{a}\dot{\mathrm{a}}^{2}\mathbf{T}^{2}f_{0},
\\\nonumber
\mathcal{I}_{2}&=& -\mathrm{a}\dot{\mathrm{a}}\varepsilon
\varphi-\mathrm{a}^{2}\varepsilon \varphi \mathbb{R}, \quad
\mathcal{I}_{3}=-\mathrm{a}\dot{\mathrm{a}}f_{0}
-c_{1}c_{2}^{-1}\mathrm{a}^{3}\varphi f_{0}.
\end{eqnarray}
Using Eqs.(\ref{12}) and (\ref{20}), we obtain the analytic solution
of scalar field and scale factor as
\begin{eqnarray}\label{41}
\varphi&=& \mathrm{a}c_{6}-\frac{6c_{3}f_{0}}{c_{2}\varepsilon},
\\\label{42}
\mathrm{a}(\mathrm{t})&=&\bigg[6(\frac{1+3\varepsilon}{\varepsilon})
(b_{2}^{2}+(b_{1}^{2}-b_{2}^{2})\sin^{2}\mathbb{B}-2b_{1}b_{2}\sin\mathbb{B}
\cos\mathbb{B})\bigg]^{\frac{1}{(2+6\varepsilon)}},
\end{eqnarray}
where
$\mathbb{B}=\frac{\sqrt{\varepsilon(1+3\varepsilon)}\mathrm{t}}{\sqrt{6}}$.
The corresponding physical parameters become
\begin{eqnarray}\label{43}
\mathcal{H}&=&{\frac{\sqrt{\varepsilon}\left[2b_{1}b_{2}\cos^{2}\mathbb{B}-
\sin\mathbb{B}\cos\mathbb{B}(b_{1}^{2}-b_{2}^{2})-b_{1}b_{2}\right]}{\sqrt
{6(1+3\varepsilon)}\left[b_{1}^{2}\cos^{2}\mathbb{B}-b_{2}^{2}\cos^{2}
\mathbb{B}+2b_{1}b_{2}\sin\mathbb{B}\cos\mathbb{B}-b_{1}^{2}\right]}},
\\\nonumber
\mathrm{q}&=&-\bigg[b_{1}^{4}+3\varepsilon
b_{1}^{4}+3b_{1}^{2}b_{2}^{2}\varepsilon+b_{1}^{4}\cos^{4}\mathbb{B}
-6b_{1}^{2}b_{2}^{2}\cos^{4}\mathbb{B}+4b_{1}^{3}b_{2}\sin\mathbb{B}
\\\nonumber &\times&
\cos^{3}\mathbb{B}-4b_{1}b_{2}^{3}\sin\mathbb{B}\cos^{3}\mathbb{B}+b_{2}^{4}\cos^{4}
\mathbb{B}-3\varepsilon b_{1}^{4}\cos^{2}\mathbb{B}-3\sin2\mathbb{B}
\\\nonumber &\times&
\varepsilon b_{1}^{3}b_{2}-3\varepsilon b_{1}
b_{2}^{3}\sin2\mathbb{B}+3\varepsilon
b_{2}^{4}\cos^{2}\mathbb{B}-2b_{1}^{4}\cos^{2}\mathbb{B}
-2b_{1}^{3}b_{2}\sin2\mathbb{B}
\\\nonumber &+&
6b_{1}^{2}b_{2}^{2}\cos^{2}\mathbb{B}\bigg]\bigg[\cos^{4}\mathbb{B}
b_{1}^{4}-6b_{1}^{2}b_{2}^{2}\cos^{4}\mathbb{B}+b_{2}^{4}\cos^{4}
\mathbb{B}+6b_{1}^{2}b_{2}^{2}\cos^{2}\mathbb{B}
\\\nonumber &+&
4b_{1}^{3}b_{2}\sin\mathbb{B}\cos^{3}\mathbb{B}-4b_{1}b_{2}^{3}
\sin\mathbb{B}\cos^{3}\mathbb{B}-b_{1}^{4}\cos^{2}\mathbb{B}-b_{2}
^{4}\cos^{2}\mathbb{B}
\\\label{44}&-&b_{1}^{3}b_{2}\sin2\mathbb{B}
+b_{1}b_{2}^{3}\sin2\mathbb{B}-b_{1}^{2}b_{2} ^{2}\bigg]^{-1}.
\end{eqnarray}
The value of $r-s$ parameters is given in appendix \textbf{B}.
\begin{figure}
\epsfig{file=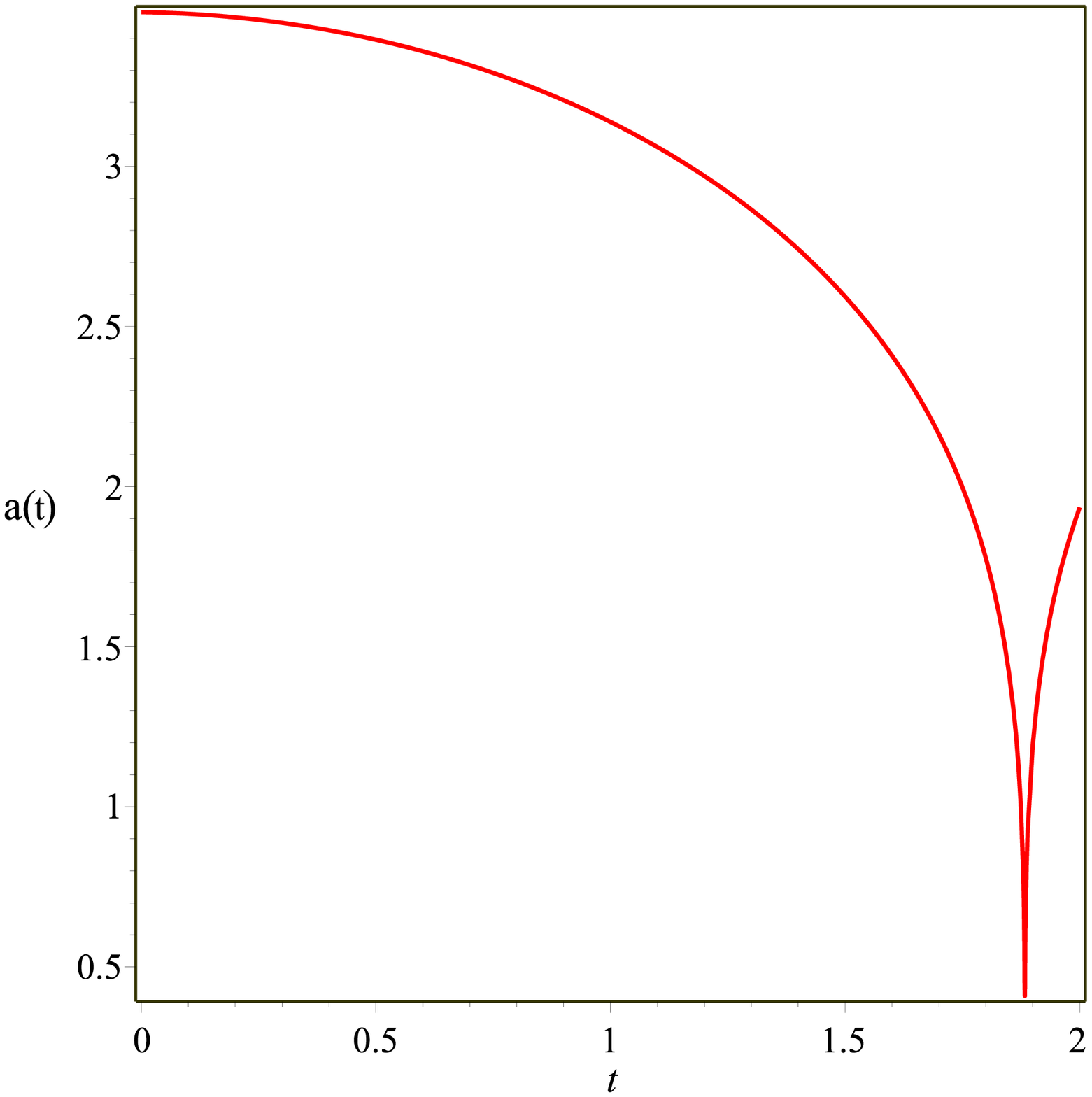,width=.5\linewidth}
\epsfig{file=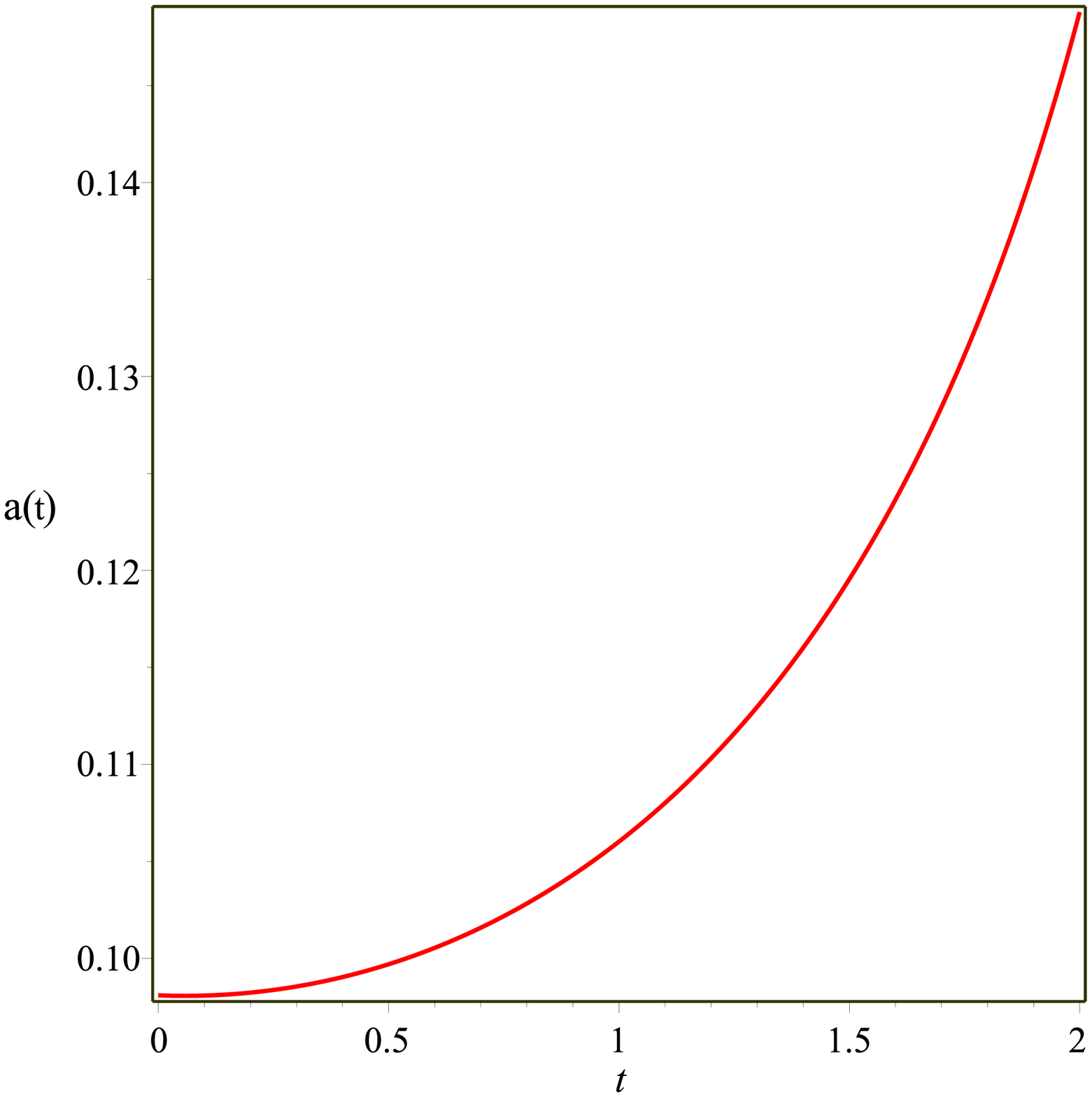,width=.5\linewidth} \caption{Graphs of scale
parameters for $\varepsilon=1$ (Left) and $\varepsilon=-1$ (Right).}
\end{figure}
\begin{figure}
\epsfig{file=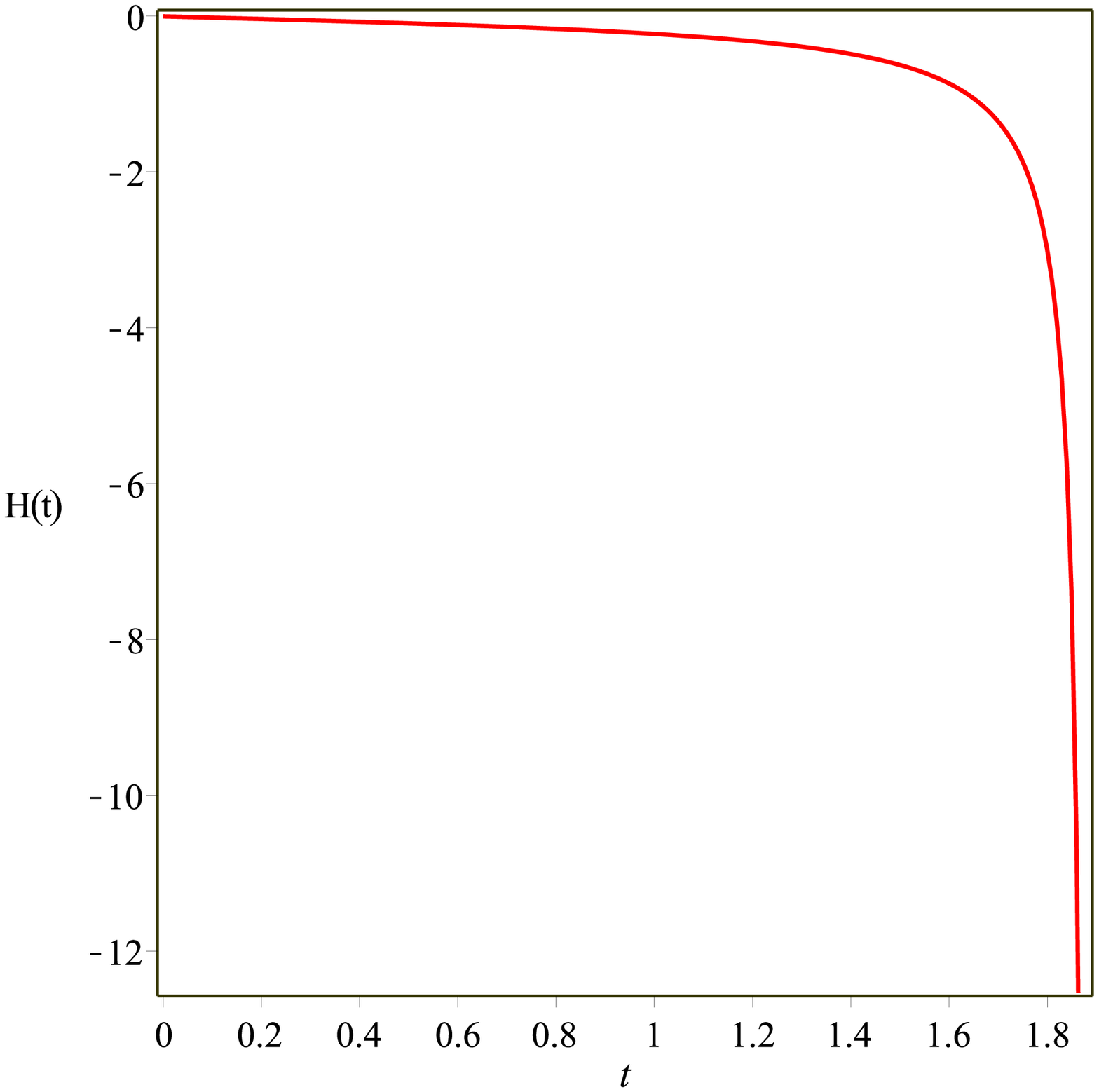,width=.5\linewidth}
\epsfig{file=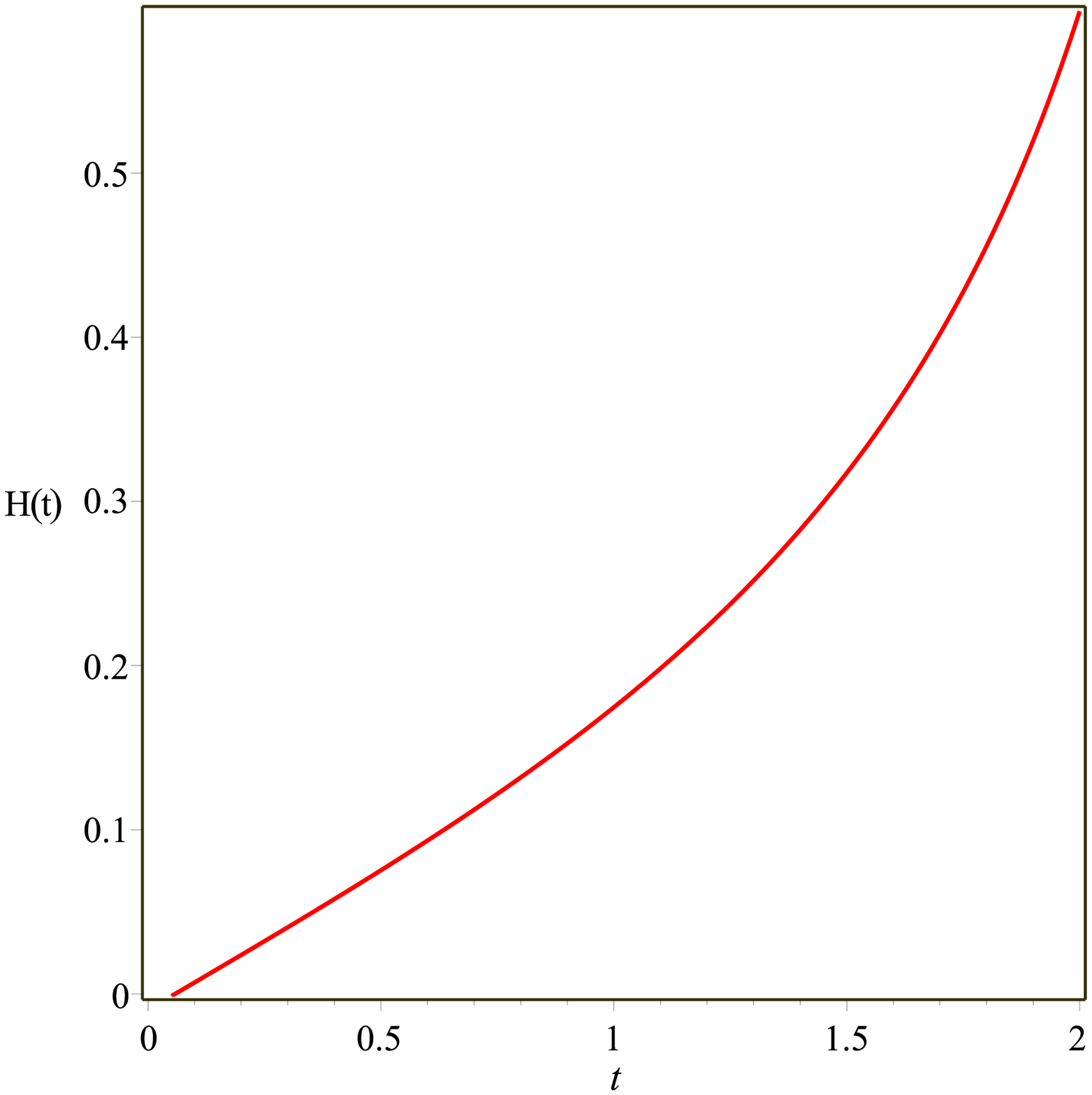,width=.5\linewidth} \caption{Behavior of Hubble
Parameter for $\varepsilon=1$ (Left) and $\varepsilon=-1$ (Right).}
\end{figure}
\begin{figure}
\epsfig{file=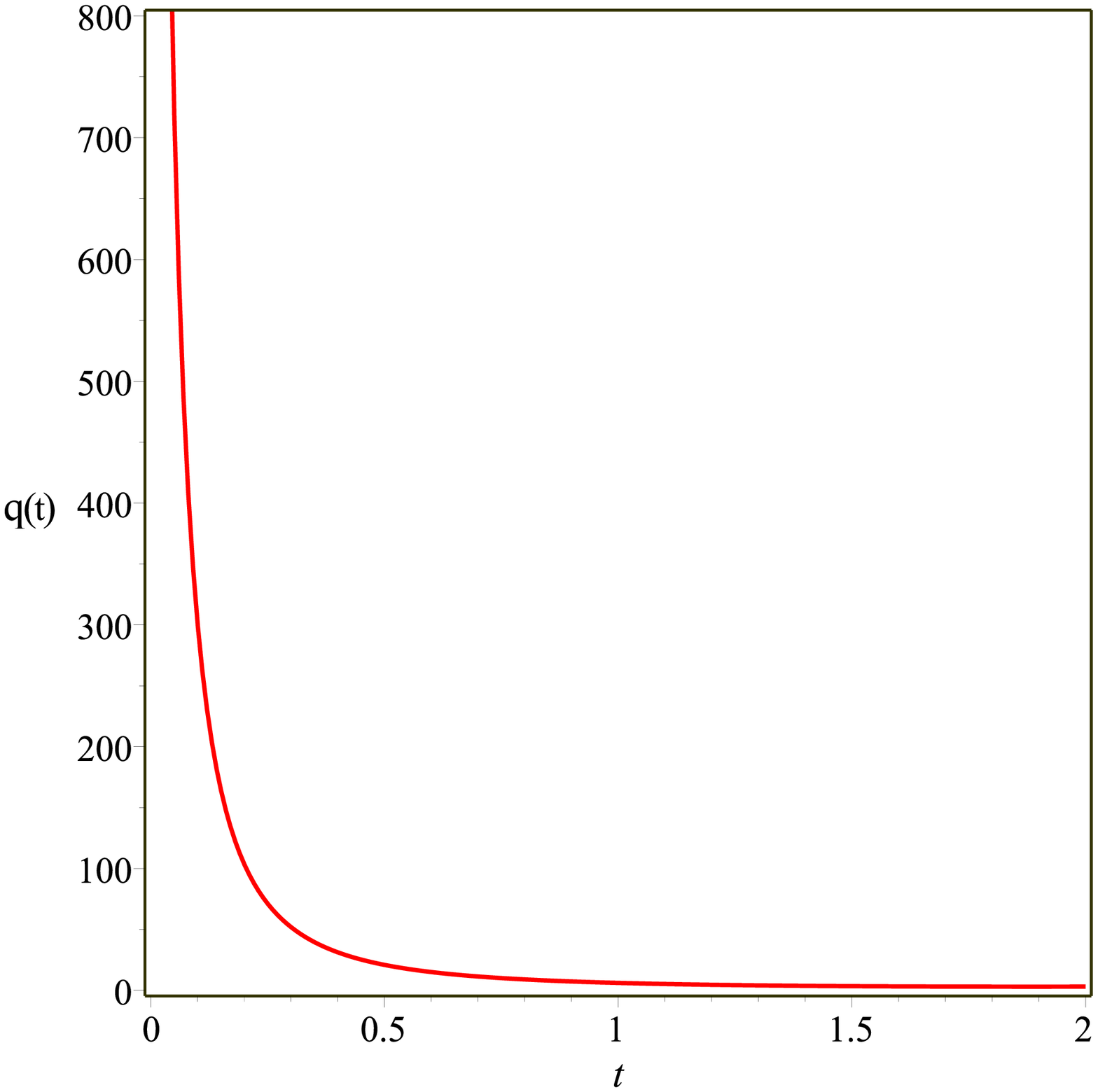,width=.5\linewidth}
\epsfig{file=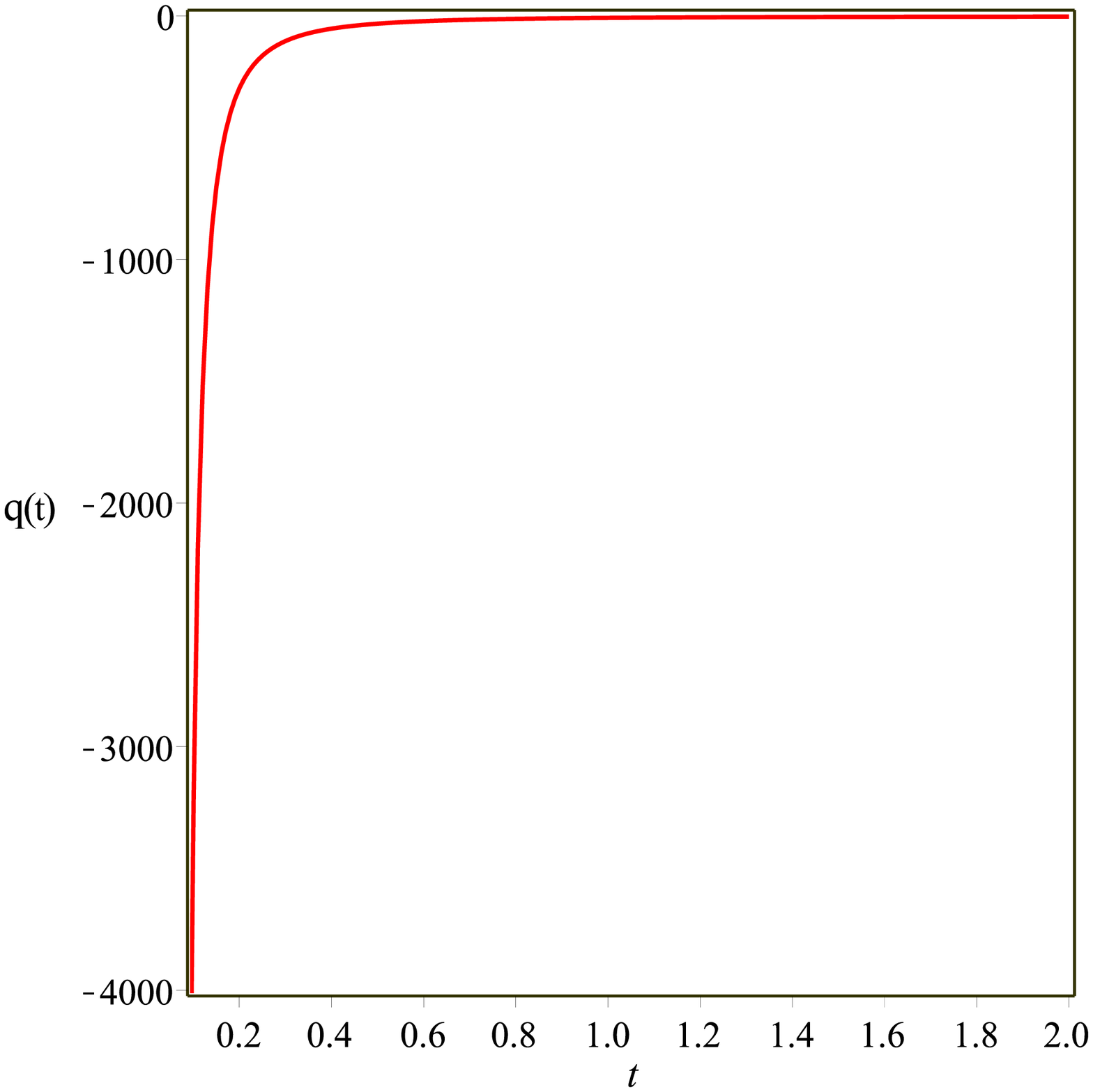,width=.5\linewidth} \caption{Deceleration
Parameter for $\varepsilon=1$ (Left) and $\varepsilon=-1$ (Right).}
\end{figure}
\begin{figure}
\epsfig{file=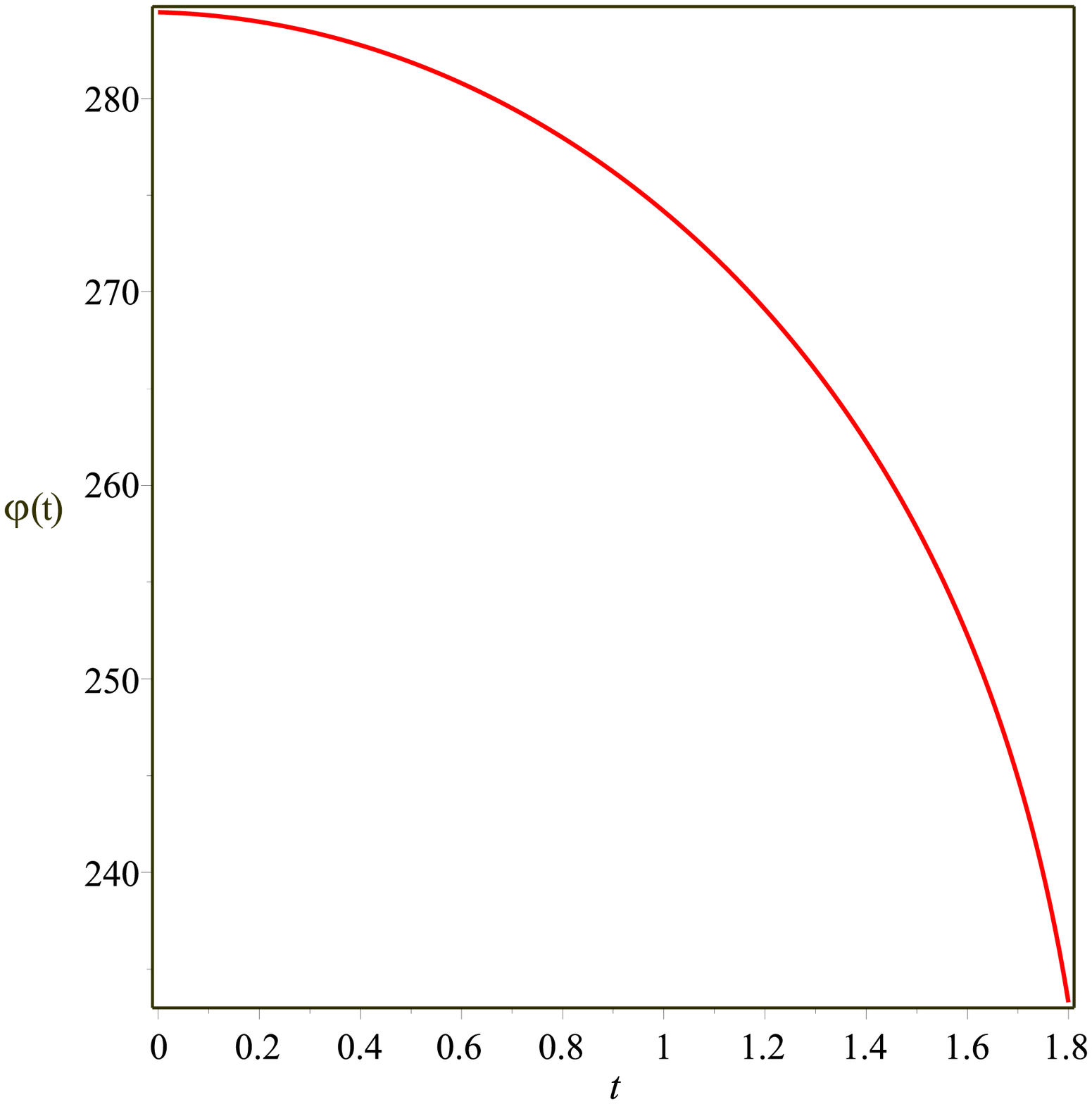,width=.5\linewidth}
\epsfig{file=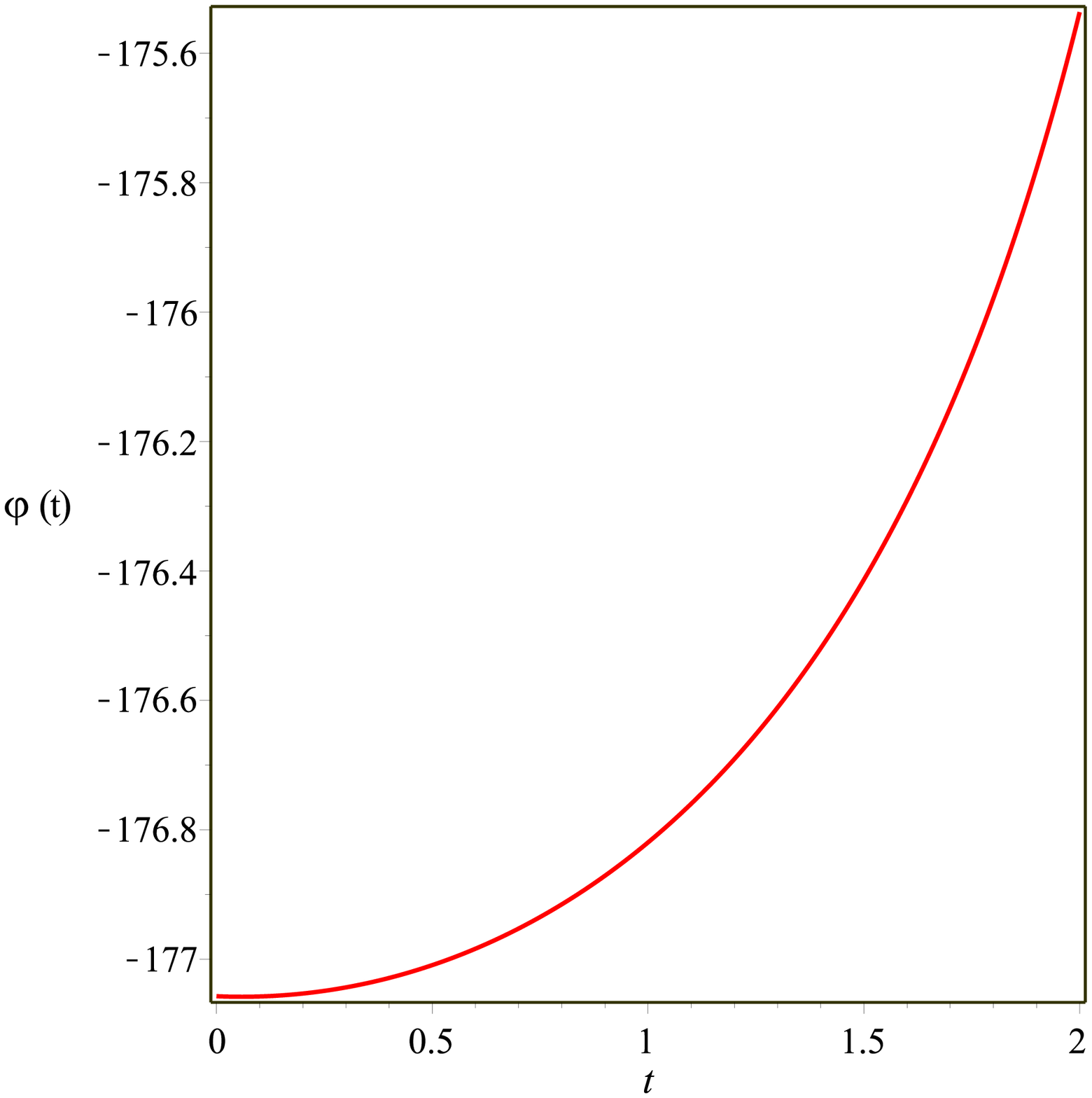,width=.5\linewidth} \caption{Behavior of scalar
field for $\varepsilon=1$ (Left) and $\varepsilon=-1$ (Right).}
\end{figure}
\begin{figure}
\epsfig{file=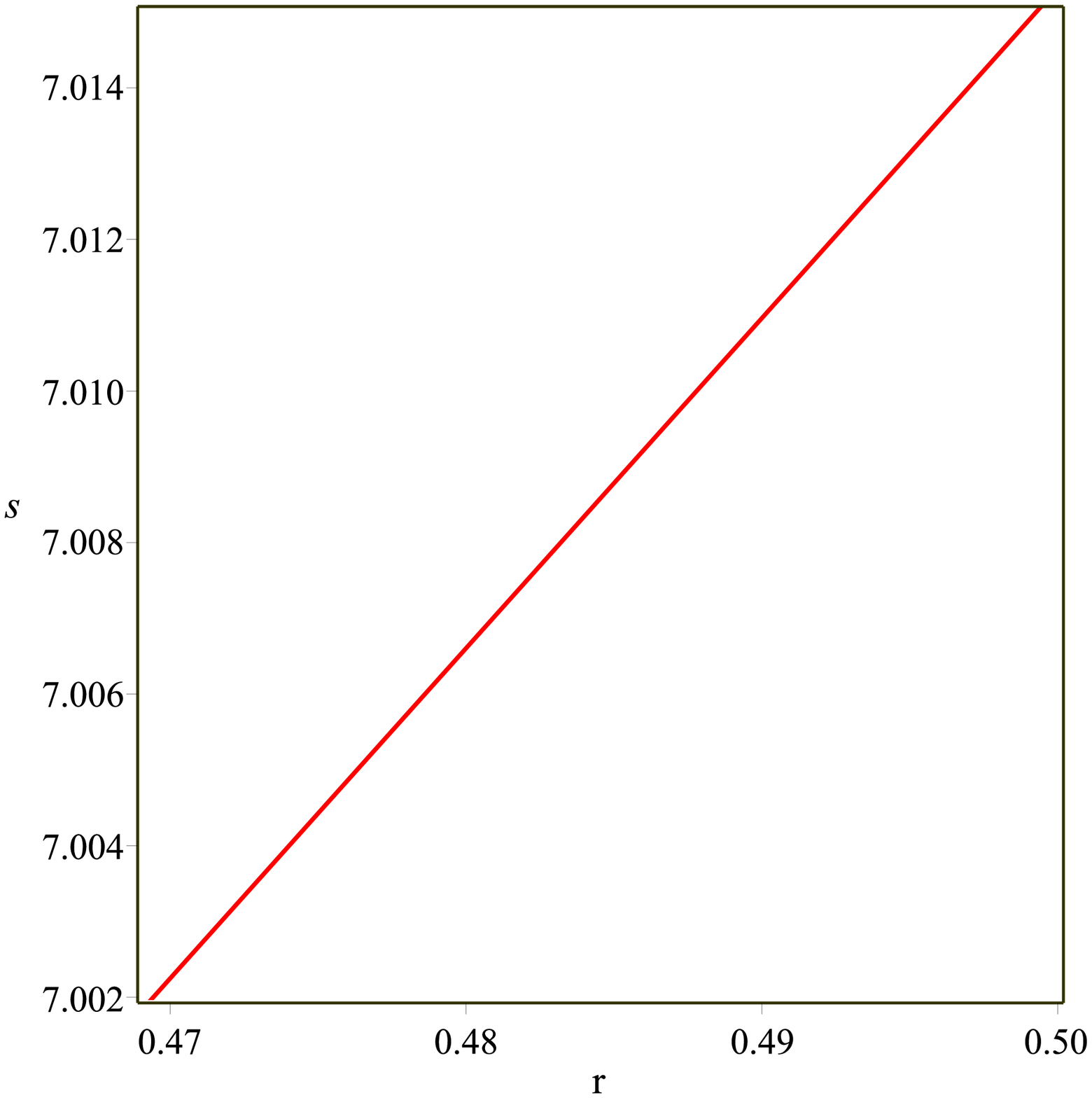,width=.5\linewidth}
\epsfig{file=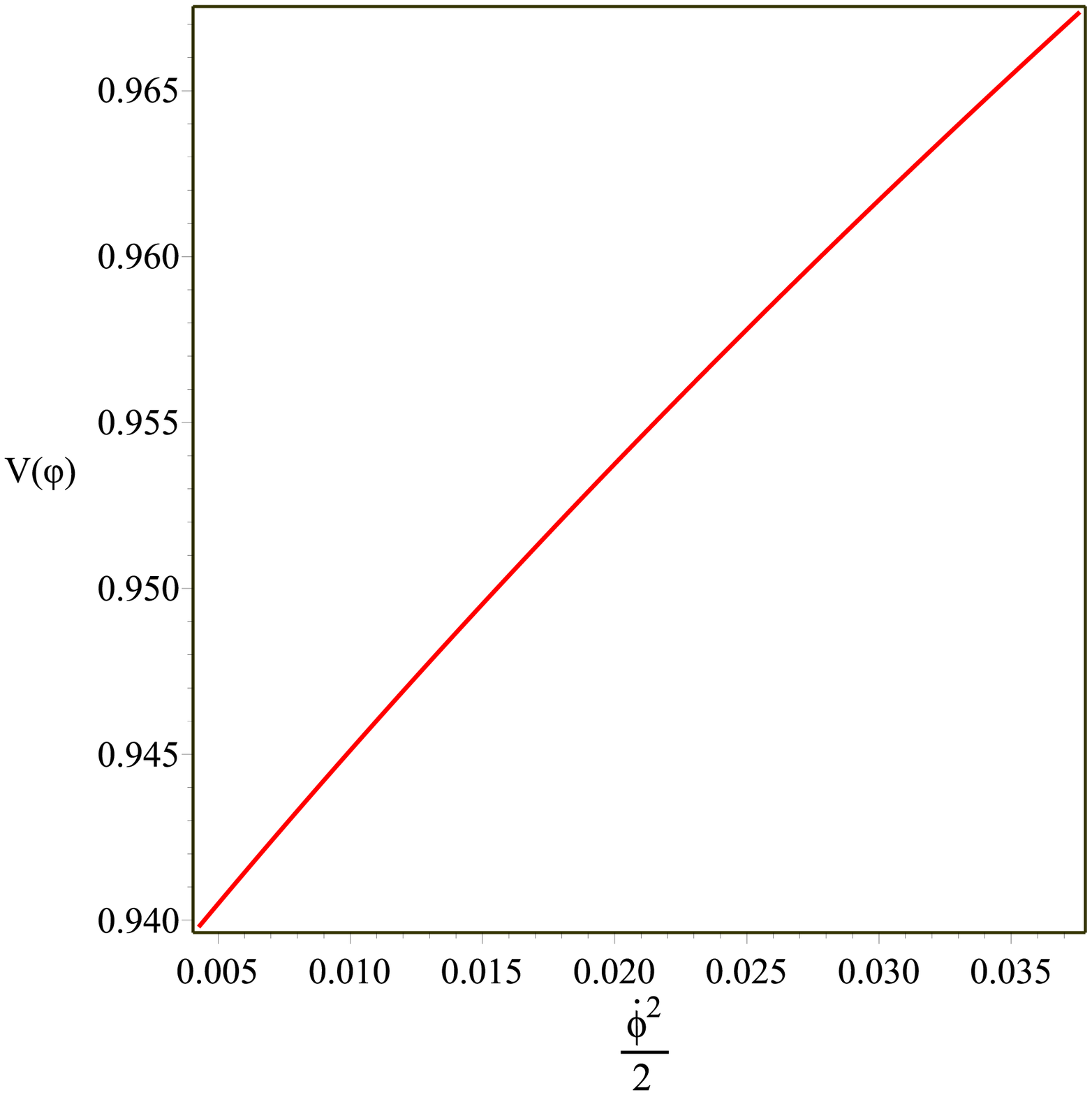,width=.5\linewidth} \caption{Graphs of $r-s$
parameter (left) and potential energy (right) for $\varepsilon=-1$}
\end{figure}

The graphical behavior of $\mathrm{a}(\mathrm{t})$ corresponding to
$\varepsilon=1,-1$ is given in Figure \textbf{6}. The left graph
determines that initially the scale factor is positively decreasing
but with time it shows increasing behavior. The right graph shows
that $\mathrm{a}(\mathrm{t})$ is positively increasing for the
phantom model which describes cosmic expansion. Figures \textbf{7}
and \textbf{8} determine the expanding behavior of the universe for
the phantom model. Figure \textbf{9} analyzes the behavior of scalar
field corresponding to quintessence and phantom models. The right
graph represents $\varphi<0$ indicating accelerated expansion while
$\varphi>0$ for the quintessence model leads to the decelerated
cosmic expansion. Figure \textbf{10} (left) shows that $r-s$
parameters support quintessence and phantom DE eras while the right
plot ensures $\frac{\dot{\varphi}^{2}}{2}<\vartheta$ implying that
phantom models yield accelerated expansion.

\section{Conclusions}

Modified theories play a crucial role to examine the dark universe
because of extra geometric terms. In this article, we have found
analytic solutions of flat FRW spacetime with the scalar field in
the background of EMSG. For this purpose, we consider the NS
approach to find the exact cosmological solutions. The Lagrangian
minimizes the complexity of the system and helps to establish
analytic solutions. We have formulated the Lagrangian in the
background of EMSG and established symmetry generators with
conserved quantities. The analytical solutions have been examined
for various models and investigated their behavior through various
cosmological quantities.

We have developed four non-zero symmetry generators and conserved
factors in the first case. We have obtained exact solutions
corresponding to quintessence and phantom models. The graphical
behavior of the scale parameter and expansion rate is found to be
increasing for both scalar field models (Figures \textbf{1} and
\textbf{2}). The deceleration parameter and scalar field remain
negative (Figures \textbf{3} and \textbf{4}). The $r-s$ parameters
give quintessence and phantom DE phases for $s>0$ and $r<1$,
respectively while the dominating potential energy shows expanding
behavior of the universe (Figure \textbf{5}). The resulting solution
in the second case implies that our cosmos is in the accelerated
expansion phase corresponding to the phantom model (Figures
\textbf{6} -\textbf{10}).

We have observed that the symmetry generators and corresponding
conserved parameters exist for both models. The maximum symmetry
generators as well as conserved parameters are obtained for the
first model. This indicates that the minimal model gives more viable
results than a non-minimal model. It is noteworthy that our
solutions are more compatible with current observational data as
compared to other modified gravitational theories. For example, in
$f(\mathbb{R},\mathcal{T})$ theory, the solutions obtained for the
quintessence model are not compatible with late-time cosmic
acceleration \cite{43}. In teleparallel theory, the only phantom
model describes the present state of the universe \cite{40}. It
would be fascinating to extend this work for anisotropic spacetime
which could provide a useful framework to study the cosmic mysteries
in EMSG.

The study of scalar field models has become the subject of great
interest for researchers due to their physical importance in
cosmology. These models support the cosmic accelerated expansion and
help to solve the horizon problem. In the early times, these models
determine rapid expansion in the inflationary era and present
acceleration in various DE models \cite{44}. The scalar field's
kinetic and potential energies are important in the study of cosmic
expansion. The potential and kinetic energies of the scalar field
are the key factors in the study of cosmic expansion. The scalar
field $(\varphi)$ must be negative and potential energy
$(\vartheta)$ should be greater than kinetic energy
$(\frac{\dot{\varphi}^{2}}{2})$ for accelerated expansion whereas
$(\frac{\dot{\varphi}^{2}}{2}>\vartheta)$ represents decelerated
expansion of the universe. We have found that scalar field supports
the accelerated expansion of the universe as it is negative and
potential energy is greater than kinetic energy in both minimal as
well as non-minimal models.

Recently, we have explored exact cosmological solutions of FRW
spacetime through the NS technique in EMSG \cite{37}. We have
considered minimal as well as non-minimal models of this theory and
obtained exact solutions through conserved quantities. We have
concluded that conserved quantities are very helpful to obtain
viable cosmological solutions. The scale factor $(\mathrm{a})$ in
both models grows continuously which indicates that the universe
experiences accelerated expansion while the Hubble $(\mathcal{H})$
and deceleration $(\mathrm{q})$ parameters identify the decreasing
rate of expansion. In the present manuscript, we have generalized
this work with the minimal coupling of scalar field and matter in
the context of EMSG and obtained exact solutions through the
Euler-Lagrange equations of motion. Here, all the physical
parameters $(\mathrm{a},\mathcal{H},\mathrm{q})$ and scalar field
support the accelerated expansion of the universe in both minimal as
well as non-minimal models. Hence, the addition of a scalar field
affects the behavior of all physical quantities such that all
parameters determine the current cosmic accelerated expansion.


\section*{Appendix A}
\renewcommand{\theequation}{A\arabic{equation}}
\setcounter{equation}{0}

The values of $r-s$ parameters corresponding to the minimal coupling
model are
\begin{eqnarray}\nonumber
r&=&\bigg[18{{\mathrm{a}_{{2}}}^{4}\varepsilon}^{2}+18a_{1}^{2}a_{2}^{2}
\varepsilon^{2}+3{\mathrm{a}_{{1}}}^{2}{\mathrm{a}_{{2}}}^{2}\varepsilon
+3{\mathrm{a}_{{2}}}^{4}\varepsilon-{\mathrm{a}_{{1}}}^{2}{\mathrm{a}_{
{2}}}^{2}+4a_{1}^{3}a_{2}\sin\mathbb{A}\cos^{3}\mathbb{A}\\\nonumber
&-&
6a_{1}^{2}a_{2}^{2}\cos^{4}\mathbb{A}+\mathrm{a}_{1}^{4}\cos^{4}\mathbb
{A}+a_{2}^{4}\cos^{4}\mathbb{A}-4a_{1}a_{2}^{3}\sin\mathbb{A}\cos^{3}
\mathbb{A}+18\cos^{2}\mathbb{A} \\\nonumber &\times&
a_{1}^{4}\varepsilon^{2}-18a_{2}^{4}\varepsilon^{2}\cos^{2}\mathbb{A}
+36a_{1}^{3}a_{2}\varepsilon^{2}\sin\mathbb{A}\cos\mathbb{A}
+2\mathrm{a}_{{1}}{\mathrm{a}_{{2}}}^{3}\sin\mathbb{A}\cos\mathbb{A}+a_{1}
\\\nonumber &\times&
36\varepsilon^{2}a_{2}^{3}\sin\mathbb{A}\cos\mathbb{A}+3a_{1}^{4}\varepsilon
\cos^{2}\mathbb{A}-3a_{2}^{4}\varepsilon\cos^{2}\mathbb{A}+6a_{1}^{2}a_{2}^{2}
\cos^{2}\mathbb{A}+\sin2\mathbb{A}
\\\nonumber &\times&
3a_{1}^{3}a_{2}\varepsilon+3a_{1}a_{2}^{3}\varepsilon\sin2\mathbb
{A}-a_{1}^{4}\cos^{2}\mathbb{A}-a_{2}^{4}\cos^{2}\mathbb{A}
-2a_{1}^{3}\cos\mathbb{A}a_{2}\sin\mathbb{A}\bigg]
\\\nonumber &\times&
\bigg[a_{1}^{4}\cos^{4}\mathbb{A}-6a_{1}^{2}a_{2}^{2}
\cos^{4}\mathbb{A}+4a_{1}^{3}a_{2}\sin\mathbb{A}\cos^{3}\mathbb{A}-4a_{1}a_{2}
^{3}\sin\mathbb{A}\cos^{3}\mathbb{A}
\\\nonumber &+&
a_{2}^{4}\cos^{4}\mathbb{A}+6a_{1}^{2}a_{2}^{2}\cos^{2}
\mathbb{A}+4a_{1}a_{2}^{3}\sin\mathbb{A}\cos\mathbb{A}-2a_{2}^{4}\cos^{2}
\mathbb{A}+a_{2}^{4}\bigg]^{-1},
\\\nonumber
s&=&-2/3\bigg[-a_{2}^{6}-a_{1}^{2}a_{2}^{4}+3\varepsilon
a_{2}^{6}+3a_{1}^{2}a_{2}^{4}\varepsilon+18a_{2}^{6}\varepsilon^{2}
+18a_{1}^{2}a_{2}^{4}\varepsilon^{2}-90\cos^{4}\mathbb{A}
\\\nonumber&\times&
\mathrm{a}_{1
}^{2}a_{2}^{4}\varepsilon^{2}+12a_{1}a_{2}^{5}\varepsilon\sin\mathbb{A}
\cos\mathbb{A}+72a_{1}^{5}a_{2}\varepsilon^{2}\sin\mathbb{A}\cos^{3}
\mathbb{A}-90a_{1}^{4}a_{2}^{2}\cos^{4}\mathbb{A}
\\\nonumber &\times&
\varepsilon^{2}-72a_{1}a_{2}^{5}\varepsilon^{2}\sin\mathbb{A}\cos^{3}
\mathbb{A}+12a_{1}^{5}a_{2}\varepsilon\sin\mathbb{A}\cos^{3}\mathbb{A}
+72a_{1}^{3}\sin2\mathbb{A}\cos\mathbb{A}
\\\nonumber &\times&
\varepsilon^{2}a_{2}^{3}-12a_{1}a_{2}^{5}\varepsilon\sin^{3}\mathbb{A}
\cos\mathbb{A}+36a_{1}a_{2}^{5}\varepsilon^{2}\sin2\mathbb{A}
+6a_{1}^{3}a_{2}^{3}\varepsilon\sin2\mathbb{A}-a_{1}^{4}a_{2}^{2}
\\\nonumber &\times&
15\varepsilon\cos^{4}\mathbb{A}
-15a_{1}^{2}a_{2}^{4}\varepsilon\cos^{4}\mathbb{A}-4a_{1}^{5}a_{2}\sin
\mathbb{A}\cos^{3}\mathbb{A}+4a_{1}a_{2}^{5}\sin\mathbb{A}\cos^{3}\mathbb{A}
\\\nonumber &+&
108a_{1}^{4}a_{2}^{2}
\varepsilon^{2}\cos\mathbb{A}^{2}-4a_{1}^{3}a_{2}^{3}\sin\mathbb{A}
\cos\mathbb{A}-4a_{1}\cos\mathbb{A}a_{2}^{5}\sin\mathbb{A}
+72\cos^{2} \mathbb{A}
\\\nonumber &\times&
\varepsilon^{2}a_{1}^{2}a_{2}^{4}+12
a_{1}^{2}a_{2}^{4}\varepsilon\cos^{2}\mathbb{A}
+18a_{1}^{4}\mathrm{a}_{ 2}^{2}\varepsilon\cos^{2}\mathbb{A}+
18a_{1}^{6}\varepsilon^{2}\cos^{4}\mathbb{A}+3\cos^{4}\mathbb{A}
\\\nonumber &\times&
a_{1}^{6}\varepsilon+5a_{1}^{4}a_{2}^{2}\cos^{4}\mathbb{A}+3a_{2}^{6}
\varepsilon\cos^{4}\mathbb{A}+18a_{2}^{6}\varepsilon^{2}\cos^{4}\mathbb{A}
+5a_{1}^{2}a_{2}^{4}\cos^{4}\mathbb{A}-\varepsilon^{2}
\\\nonumber &\times&
36a_{2}^{6}\cos^{2}\mathbb{A}-6a_{1}^{4}a_{2}^{2}
\cos^{2}\mathbb{A}-4a_{1}^{2}\cos^{2}\mathbb{A}a_{2}^{4}-6a_{2}^{6}
\varepsilon\cos^{2}\mathbb{A}+2a_{2}^{6}\cos^{2}
\mathbb{A}
\\\nonumber &-&
a_{1}^{6}\cos^{4}\mathbb{A}-a_{2}^{6}\cos^{4}
\mathbb{A}\bigg]\bigg[a_{2}^{6}-2a_{1}^{2}a_{2}^{4}
6a_{2}^{6}\varepsilon-6a_{1}^{2}a_{2}^{4}\varepsilon-6a_{2}^{6}\varepsilon
\cos^{4}\mathbb{A}-a_{1}^{3}a_{2}^{3}
\\\nonumber &\times&
12\varepsilon\sin2\mathbb{A}
-24a_{1}a_{2}^{5}\varepsilon\sin\mathbb{A}+
24a_{2}^{5}\varepsilon\sin\mathbb{A}\cos^{3}\mathbb{A}
-24a_{1}^{5}\varepsilon\sin\mathbb{A}\cos^{3}\mathbb{A} a_{2}
\\\nonumber &+&
30a_{1}^{4}a_{2}^{2}\varepsilon\cos^{4}\mathbb{A}-8a_{1}^{5}a_{2}
\sin\mathbb{A}\cos^{3}\mathbb{A}+18a_{1}^{5}\mathrm{a}
_{2}\sin\mathbb{A}\cos^{5}\mathbb{A}+30
\varepsilon\cos^{4}\mathbb{A}
\\\nonumber &\times&
a_{1}^{2}a_{2}^{4}-\cos^{4}\mathbb{A}-28a_{1}a_{2}^{5}\sin\mathbb{A}\cos^{3}\mathbb{A}-36
a_{1}^{4}a_{2}^{2}\varepsilon\cos^{2}\mathbb{A}-8a_{1}^{3}\sin\mathbb{A}\cos\mathbb{A}
\\\nonumber &\times&
a_{2}
^{3}-24a_{1}^{2}a_{2}^{4}\varepsilon\cos^{2}\mathbb{A}+10a_{1}a_{2}^{5}\sin\mathbb{A}
\cos\mathbb{A}+18\mathrm{a}_{2}^{5}a_{1}\sin\mathbb{A}\cos^{5}\mathbb{A}
-6\mathrm{a}_{1}^{6}\varepsilon
\\\nonumber &\times&
\cos^{4}\mathbb{A}-60a_{1}^{3}a_{2}^{3}\sin\mathbb{A}+60a_{1}^{3}a_{2}
^{3}\sin\mathbb{A}\cos^{3}\mathbb{A}+3a_{1}^{6}\cos^{6}\mathbb{A}
-3a_{2}^{6}\cos^{6}\mathbb{A}
\\\nonumber &+&
55a_{1}^{4}a_{2}^{2}\cos^{4}\mathbb{A}-80a_{1}^{2}a_{2}^{4}\cos^{4}\mathbb{A}
+12a_{2}^{6}\varepsilon\cos^{2}\mathbb{A}+7a_{2}^{6}\cos^{4}\mathbb{A}
-\cos^{2}\mathbb{A}
\\\nonumber &\times&
5a_{2}^{6}-45a_{1}^{4}a_{2}^{2}\cos^{6}\mathbb{A}
+45a_{1}^{2}a_{2}^{4}\cos^{6}\mathbb{A} + 37a_{1}^{2}
a_{2}^{4}-12a_{1}^{4}a_{2}^{2}\cos^{2}\mathbb{A}\bigg]^{-1}.
\end{eqnarray}

\section*{Appendix B}
\renewcommand{\theequation}{A\arabic{equation}}
\setcounter{equation}{0}

 The values of $r-s$ parameters for the non-minimal coupling model are
\begin{eqnarray}\nonumber
r&=&\bigg[486b_{1}^{8}b_{2}^{2}\varepsilon^{2}\sin\mathbb{B}
\cos^{7}\mathbb{B}+486b_{1}^{2}b_{2}^{8}\varepsilon^{2}\sin
\mathbb{B}\cos^{7}\mathbb{B}+90b_{1}^{4}b_{2}^{6}
\varepsilon^{2}\sin\mathbb{B}\cos\mathbb{B}
\\\nonumber
&-&9b_{1}^{8}b_{2}^{2}\varepsilon\sin\mathbb{B}\cos\mathbb{B}
+6b_{1}^{6}b_{2}^{4}\varepsilon\sin\mathbb{B}\cos\mathbb{B}
+15b_{1}^{4}b_{2}^{6}\varepsilon\sin\mathbb{B}\cos\mathbb{B}
-141b_{1}^{8}b_{2}^{2}\varepsilon
\\\nonumber
&\times&\sin\mathbb{B}\cos^{5}\mathbb{B}+168b_{1}^{6}b_{2}^{4}
\varepsilon\sin\mathbb{B}\cos^{5}\mathbb{B}+210b_{1}^{4}b_{2}
^{6}\varepsilon\sin\mathbb{B}\cos^{5}\mathbb{B}-102b_{1}^{2}
b_{2}^{8}\varepsilon
\\\nonumber
&\times&\sin\mathbb{B}\cos^{5}\mathbb{B}+414b_{1}^{8}b_{2}^{2}
\varepsilon^{2}\sin\mathbb{B}\cos^{3}\mathbb{B}378b_{1}^{6}
b_{2}^{4}\varepsilon^{2}\sin\mathbb{B}\cos^{3}\mathbb{B}
-630b_{1}^{4}b_{2}^{6}\varepsilon^{2}
\\\nonumber
&\times&\sin\mathbb{B}\cos^{3}\mathbb{B}+180b_{1}^{2}b_{2}^{8}
\varepsilon^{2}\sin\mathbb{B}\cos^{3}\mathbb{B}+69b_{1}^{8}b_{2}
^{2}\varepsilon\sin\mathbb{B}\cos^{3}\mathbb{B}-63b_{1}^{6}b_{2}
^{4}\varepsilon
\\\nonumber
&\times&\sin\mathbb{B}\cos^{3}\mathbb{B}-105b_{1}^{4}b_{2}^{6}
\varepsilon\sin\mathbb{B}\cos^{3}\mathbb{B}+30b_{1}^{2}b_{2}^{8}
\varepsilon\sin\mathbb{B}\cos^{3}\mathbb{B}-54b_{1}^{8}b_{2}^{2}
\varepsilon^{2}
\\\nonumber
&\times&\sin\mathbb{B}\cos\mathbb{B}+36b_{1}^{6}b_{2}^{4}
\varepsilon^{2}\sin\mathbb{B}\cos\mathbb{B}+81b_{1}^{8}
b_{2}^{2}\varepsilon\sin\mathbb{B}\cos^{7}\mathbb{B}
-126b_{1}^{6}\sin\mathbb{B}
\\\nonumber
&\times&b_{2}^{4}\varepsilon\cos^{7}\mathbb{B}-126b_{1}^{4}b_{2}
^{6}\varepsilon\sin\mathbb{B}\cos^{7}\mathbb{B}+81b_{1}^{2}
b_{2}^{8}\varepsilon\sin\mathbb{B}\cos^{7}\mathbb{B}-846b_{1}
^{8}\sin\mathbb{B}
\\\nonumber
&\times&b_{2}^{2}\varepsilon^{2}\cos^{5}\mathbb{B}+1008b_{1}^{6}
b_{2}^{4}\varepsilon^{2}\sin\mathbb{B}\cos^{5}\mathbb{B}+1260
b_{1}^{4}b_{2}^{6}\varepsilon^{2}\sin\mathbb{B}\cos^{5}
\mathbb{B}-\sin\mathbb{B}
\\\nonumber
&\times&612b_{1}^{2}b_{2}^{8}\varepsilon^{2}\cos^{5}\mathbb{B}
-756b_{1}^{6}b_{2}^{4}\varepsilon^{2}\sin\mathbb{B}\cos^{7}
\mathbb{B}--b_{1}^{10}\sin\mathbb{B}\cos^{9}\mathbb{B}
+\varepsilon^{2}\sin\mathbb{B}
\\\nonumber
&\times&756b_{1}^{4}
b_{2}^{6}\cos^{7}\mathbb{B}+b_{2}^{10}\sin\mathbb{B}\cos^{9}
\mathbb{B}-b_{2}^{10}\sin\mathbb{B}\cos^{7}\mathbb{B}+102
b_{1}^{8}b_{2}^{2}\cos^{5}\mathbb{B}\sin\mathbb{B}
\\\nonumber
&+&b_{1}^{10}\sin\mathbb{B}\cos^{3}\mathbb{B}-336b_{1}^{6}
b_{2}^{4}\sin\mathbb{B}\cos^{5}\mathbb{B}+210b_{1}^{4}b_{2}
^{6}\sin\mathbb{B}\cos^{5}\mathbb{B}-21b_{1}^{2}\sin\mathbb{B}
\\\nonumber
&\times&b_{2}^{8}\cos^{5}\mathbb{B}+252b_{1}^{9}b_{2}\varepsilon
^{2}\cos^{4}\mathbb{B}-1206b_{1}^{7}b_{2}^{3}\varepsilon^{2}
\cos^{4}\mathbb{B}-378b_{1}^{5}b_{2}^{5}\varepsilon^{2}\cos^{4}
\mathbb{B}+b_{1}^{3}
\\\nonumber
&+&990b_{2}^{7}\varepsilon^{2}\cos^{4}\mathbb{B}-90b_{1}b_{2}^{9}
\varepsilon^{2}\cos^{4}\mathbb{B}-18b_{1}^{10}\varepsilon^{2}
\sin\mathbb{B}\cos^{3}\mathbb{B}+42\varepsilon b_{1}^{9}b_{2}
\cos^{4}\mathbb{B}
\\\nonumber
&-&201b_{1}^{7}b_{2}^{3}\varepsilon\cos^{4}\mathbb{B}
-63b_{1}^{5}b_{2}^{5}\varepsilon\cos^{4}\mathbb{B}
+165b_{1}^{3}b_{2}^{7}\varepsilon\cos^{4}\mathbb{B}
-15b_{1}b_{2}^{9}\varepsilon\cos^{4}\mathbb{B}-3\varepsilon
\\\nonumber
&\times&b_{1}^{10}\sin\mathbb{B}\cos^{3}\mathbb{B}
-33b_{1}^{8}b_{2}^{2}\sin\mathbb{B}\cos^{3}\mathbb{B}
+91b_{1}^{6}b_{2}^{4}\sin\mathbb{B}\cos^{3}\mathbb{B}
-35b_{1}^{4}b_{2}^{6}\sin\mathbb{B}
\\\nonumber
&\times&\cos^{3}\mathbb{B}-54b_{1}^{9}b_{2}\varepsilon^{2}
\cos^{2}\mathbb{B}+288b_{1}^{7}b_{2}^{3}\varepsilon^{2}
\cos^{2}\mathbb{B}-180b_{1}^{3}b_{2}^{7}\varepsilon^{2}
\cos^{2}\mathbb{B}-\cos^{2}\mathbb{B}
\\\nonumber
&\times&9b_{1}^{9}b_{2}\varepsilon+27b_{1}^{5}b_{2}^{5}
\varepsilon\cos^{2}\mathbb{B}-30b_{1}^{3}b_{2}^{7}
\varepsilon\cos^{2}\mathbb{B}+3b_{1}^{8}b_{2}^{2}
\sin\mathbb{B}\cos\mathbb{B}-7b_{1}^{6}b_{2}^{4} \sin\mathbb{B}
\\\nonumber
&\times&\cos\mathbb{B}+45\cos^{9}\mathbb{B}\sin\mathbb{B}
b_{1}^{8}b_{2}^{2}-210b_{1}^{6}b_{2}^{4}\cos^{9}\mathbb{B}
\sin\mathbb{B}+210b_{1}^{4}b_{2}^{6}\cos^{9}\mathbb{B}
\sin\mathbb{B}
\\\nonumber
&-&45b_{1}^{2}b_{2}^{8}\cos^{9}\mathbb{B}\sin\mathbb{B}
+24b_{1}^{9}b_{2}\varepsilon\cos^{8}\mathbb{B}
-144b_{1}^{7}b_{2}^{3}\varepsilon\cos^{8}\mathbb{B}
+144b_{1}^{3}b_{2}^{7}\varepsilon\cos^{8}\mathbb{B}
\\\nonumber
&-&24b_{1}b_{2}^{9}\varepsilon\cos^{8}\mathbb{B}
-3b_{1}^{10}\varepsilon\sin\mathbb{B}\cos^{7}\mathbb{B}
-3b_{2}^{10}\varepsilon\sin\mathbb{B}\cos^{7}\mathbb{B}
-117b_{1}^{8}b_{2}^{2}\cos^{7}\mathbb{B}
\\\nonumber
&\times&\sin\mathbb{B}+462b_{2}^{4}\sin\mathbb{B}\cos
\mathbb{B}^{7}b_{1}^{6}-378b_{1}^{4}b_{2}^{6}\sin
\mathbb{B}\cos^{7}\mathbb{B}+63b_{1}^{2}b_{2}^{8}
\sin\mathbb{B}\cos^{7}\mathbb{B}
\\\nonumber
&-&342b_{1}^{9}b_{2}\varepsilon^{2}\cos^{6}\mathbb{B}
+252b_{1}^{5}b_{2}^{5}\varepsilon^{2}\cos^{6}\mathbb{B}
+234b_{1}b_{2}^{9}\varepsilon^{2}\cos^{6}\mathbb{B}
+36b_{1}^{10}\varepsilon^{2}\cos^{5}\mathbb{B}
\\\nonumber
&\times&\sin\mathbb{B}+18b_{2}^{10}\varepsilon^{2}\sin
\mathbb{B}\cos^{5}\mathbb{B}-57b_{1}^{9}b_{2}\varepsilon
\cos^{6}\mathbb{B}+300b_{1}^{7}b_{2}^{3}\varepsilon
\cos^{6}\mathbb{B}+\cos^{6}\mathbb{B}
\\\nonumber
&\times&42b_{1}^{5}b_{2}^{5}\varepsilon-276b_{1}^{3}b_{2}^{7}
\varepsilon\cos^{6}\mathbb{B}+39b_{1}b_{2}^{9}\varepsilon
\cos^{6}\mathbb{B}+6b_{1}^{10}\varepsilon\sin\mathbb{B}
\cos^{5}\mathbb{B}+\cos^{5}\mathbb{B}
\\\nonumber
&\times&3b_{2}^{10}\varepsilon\sin\mathbb{B}+144b_{1}^{9}b_{2}
\varepsilon^{2}\cos^{8}\mathbb{B}-864b_{1}^{7}b_{2}^{3}
\varepsilon^{2}\cos^{8}\mathbb{B}+864b_{1}^{3}b_{2}^{7}
\varepsilon^{2}\cos^{8}\mathbb{B}-b_{1}
\\\nonumber
&\times&144b_{2}^{9}\varepsilon^{2}\cos^{8}\mathbb{B}
-18b_{1}^{10}\varepsilon^{2}\sin\mathbb{B}\cos^{7}\mathbb{B}
-18b_{2}^{10}\varepsilon^{2}\sin\mathbb{B}\cos^{7}\mathbb{B}
+1800\cos^{6}\mathbb{B}
\\\nonumber
&\times&b_{1}^{7}b_{2}^{3}\varepsilon^{2}-1656b_{1}^{3}b_{2}^{7}
\varepsilon^{2}\cos^{6}\mathbb{B}+48b_{1}^{7}b_{2}^{3}
\varepsilon\cos^{2}\mathbb{B}+162b_{1}^{5}b_{2}^{5}
\varepsilon^{2}\cos^{2}\mathbb{B}-\cos^{10}\mathbb{B}
\\\nonumber
&\times&120b_{1}^{7}b_{2}^{3}-3b_{1}^{7}b_{2}^{3}\varepsilon
-18b{1}^{7}b_{2}^{3}\varepsilon^{2}-3b_{1}^{5}b_{2}^{5}
\varepsilon+10b_{1}^{9}b_{2}\cos^{10}\mathbb{B}
+252b_{1}^{5}b_{2}^{5}\cos^{10}\mathbb{B}
\\\nonumber
&-&120^{10}b_{1}^{3}b_{2}^{7}\cos\mathbb{B}-18b_{1}^{5}b_{2}^{5}
\varepsilon^{2}+10b_{1}b_{2}^{9}\cos^{10}\mathbb{B}-33b_{1}^{9}
b_{2}\cos^{8}\mathbb{B}+348\cos^{8}\mathbb{B}
\\\nonumber
&\times&b_{1}^{7}b_{2}^{3}-630b_{1}^{5}b_{2}^{5}\cos^{8}\mathbb{B}
+252b_{1}^{3}b_{2}^{7}\cos^{8}\mathbb{B}-17b_{1}b_{2}^{9}\cos^{8}
\mathbb{B}+3b_{1}^{10}\sin\mathbb{B}\cos^{7}\mathbb{B}
\\\nonumber
&+&39b_{1}^{9}b_{2}\cos^{6}\mathbb{B}-360b_{1}^{7}b_{2}^{3}
\cos^{6}\mathbb{B}+546b_{1}^{5}b_{2}^{5}\cos^{6}\mathbb{B}
-168b_{1}^{3}b_{2}^{7}\cos^{6}\mathbb{B}+\cos^{6}\mathbb{B}
\\\nonumber
&\times&7b_{1}b_{2}^{9}-3b_{1}^{10}\sin\mathbb{B}\cos^{5}
\mathbb{B}-19b_{1}^{9}b_{2}\cos^{4}\mathbb{B}+157b_{1}^{7}
b_{2}^{3}\cos^{4}\mathbb{B}-189b_{1}^{5}b_{2}^{5}\cos^{4} \mathbb{B}
\\\nonumber
&+&35b_{1}^{3}b_{2}^{7}\cos^{4}\mathbb{B}+3b_{1}^{9}b_{2}
\cos^{2}\mathbb{B}-26b_{1}^{7}b_{2}^{3}\cos^{2}\mathbb{B}
+21b_{1}^{5}b_{2}^{5}\cos^{2}\mathbb{B}+b_{1}^{7}b_{2}^{3}
\bigg]\bigg[-b_{1}^{10}
\\\nonumber
&\times&\sin\mathbb{B}\cos^{9}\mathbb{B}+b_{2}^{10}\sin
\mathbb{B}\cos^{9}\mathbb{B}-2b_{2}^{10}\sin\mathbb{B}
\cos^{7}\mathbb{B}+b_{2}^{10}\sin\mathbb{B}\cos^{5}
\mathbb{B}+\cos^{5}\mathbb{B}
\\\nonumber
&\times&55b_{1}^{8}b_{2}^{2}\sin\mathbb{B}-280b_{1}^{6}
b_{2}^{4}\sin\mathbb{B}\cos^{5}\mathbb{B}+280b_{1}^{4}
b_{2}^{6}\sin\mathbb{B}\cos^{5}\mathbb{B}-55b_{1}^{2}
\sin\mathbb{B}\cos^{5}\mathbb{B}
\\\nonumber
&\times&b_{2}^{8}-10b_{1}^{8}b_{2}^{2}\sin\mathbb{B}
\cos^{3}\mathbb{B}+70b_{1}^{6}b_{2}^{4}\sin\mathbb{B}
\cos^{3}\mathbb{B}-70b_{1}^{4}b_{2}^{6}\sin\mathbb{B}
\cos^{3}\mathbb{B}+\cos^{9}\mathbb{B}
\\\nonumber
&\times&45b_{1}^{8}b_{2}^{2}\sin\mathbb{B}-5b_{1}^{6}
b_{2}^{4}\sin\mathbb{B}\cos\mathbb{B}-210b_{1}^{6}
b_{2}^{4}\sin\mathbb{B}\cos^{9}\mathbb{B}+210b_{1}^{4}
b_{2}^{6}\sin\mathbb{B}\cos^{9}\mathbb{B}
\\\nonumber
&-&45b_{1}^{2}b_{2}^{8}\sin\mathbb{B}\cos^{9}\mathbb{B}
-90b_{1}^{8}b_{2}^{2}\cos^{7}\mathbb{B}\sin\mathbb{B}
+420b_{1}^{6}b_{2}^{4}\sin\mathbb{B}\cos^{7}\mathbb{B}
-420\cos^{7}\mathbb{B}
\\\nonumber
&\times&b_{1}^{4}b_{2}^{6}\sin\mathbb{B}+90b_{1}^{2}b_{2}
^{8}\sin\mathbb{B}\cos^{7}\mathbb{B}+10b_{1}^{2}b_{2}^{8}
\sin\mathbb{B}\cos^{3}\mathbb{B}+5b_{1}^{4}b_{2}^{6}
\sin\mathbb{B}\cos\mathbb{B}+b_{1}^{9}
\\\nonumber
&\times&10b_{2}\cos^{10}\mathbb{B}-120b_{1}^{7}b_{2}^{3}
\cos^{10}\mathbb{B}+252b_{1}^{5}b_{2}^{5}\cos^{10}\mathbb{B}
-120b_{1}^{3}b_{2}^{7}\cos^{10}\mathbb{B}+\cos^{10}\mathbb{B}
\\\nonumber
&\times&10b_{1}b_{2}^{9}-25b_{1}^{9}b_{2}\cos^{8}\mathbb{B}
+300b_{1}^{7}b_{2}^{3}\cos^{8}\mathbb{B}-630b_{1}^{5}b_{2}
^{5}\cos^{8}\mathbb{B}+300b_{1}^{3}b_{2}^{7}\cos^{8}\mathbb{B}
\\\nonumber
&-&25b_{1}b_{2}^{9}\cos^{8}\mathbb{B}+2b_{1}^{10}\sin\mathbb{B}
\cos^{7}\mathbb{B}+20b_{1}^{9}b_{2}\cos^{6}\mathbb{B}-260b_{1}
^{7}b_{2}^{3}\cos^{6}\mathbb{B}+\cos^{6}\mathbb{B}
\\\nonumber
&\times&560b_{1}^{5}b_{2}^{5}-260b_{1}^{3}b_{2}^{7}\cos^{6}
\mathbb{B}+20b_{1}b_{2}^{9}\cos^{6}\mathbb{B}-b_{1}^{10}
\sin\mathbb{B}\cos^{5}\mathbb{B}-5b_{1}^{9}b_{2}\cos^{4} \mathbb{B}
\\\nonumber
&+&90b_{1}^{7}b_{2}^{3}\cos^{4}\mathbb{B}-210b_{1}^{5}b_{2}^{5}
\cos^{4}\mathbb{B}+90b_{1}^{3}b_{2}^{7}\cos^{4}\mathbb{B}
-10b_{1}^{7}b_{2}^{3}\cos^{2}\mathbb{B}+30\cos^{2}\mathbb{B}
\\\nonumber
&\times&b_{1}^{5}b_{2}^{5}-5b_{1}b_{2}^{9}\cos^{4}\mathbb{B}
-10b_{1}^{3}b_{2}^{7}\cos^{2}\mathbb{B}-b_{1}^{5}b_{2}^{5}\bigg]^{-1},
\\\nonumber
s&=&-2/3\bigg[30b_{1}^{12}b_{2}^{2}\varepsilon\sin\mathbb{B}\cos^{3}
\mathbb{B}-255b_{1}^{10}b_{2}^{4}\varepsilon\sin\mathbb{B}\cos^{3}
\mathbb{B}+135b_{1}^{8}\varepsilon\sin\mathbb{B}\cos^{3}\mathbb{B}
\\\nonumber
&\times&b_{2}^{6}+315b_{1}^{6}b_{2}^{8}\varepsilon\sin\mathbb{B}\cos
^{3}\mathbb{B}-105\sin\mathbb{B}\cos^{3}\mathbb{B}b_{1}^{4}b_{2}
^{10}\varepsilon+549b_{1}^{12}\varepsilon\sin\mathbb{B}\cos^{7}
\mathbb{B}
\\\nonumber
&\times&b_{2}^{2}-3267b_{1}^{10}b_{2}^{4}\varepsilon\sin\mathbb{B}
\cos^{7}\mathbb{B}+2475b_{1}^{8}b_{2}^{6}\varepsilon\sin\mathbb{B}
\cos^{7}\mathbb{B}+3465b_{1}^{6}b_{2}^{8}\varepsilon\cos^{7}\mathbb{B}
\\\nonumber
&\times&\sin\mathbb{B}-2541b_{1}^{4}b_{2}^{10}\varepsilon\sin\mathbb{B}
\cos^{7}\mathbb{B}+291b_{1}^{2}\varepsilon\sin\mathbb{B}b_{2}^{12}
\cos^{7}\mathbb{B}-1332b_{1}^{12}\cos^{5}\mathbb{B}
\\\nonumber
&\times&\varepsilon^{2}b_{2}^{2}\sin\mathbb{B}+8370b_{1}^{10}b_{2}^{4}
\varepsilon^{2}\sin\mathbb{B}\cos^{5}\mathbb{B}-5400b_{1}^{8}b_{2}^{6}
\varepsilon^{2}\sin\mathbb{B}\cos^{5}\mathbb{B}-\cos^{5}\mathbb{B}
\\\nonumber
&\times&9450b_{1}^{6}b_{2}^{8}\varepsilon^{2}\sin\mathbb{B}+5292b_{1}^{4}
b_{2}^{10}\varepsilon^{2}\sin\mathbb{B}\cos^{5}\mathbb{B}-378b_{2}^{12}
\varepsilon^{2}b_{1}^{2}\sin\mathbb{B}\cos^{5}\mathbb{B}
-\varepsilon
\\\nonumber
&\times&222b_{1}^{12}b_{2}^{2}\sin\mathbb{B}\cos^{5}\mathbb{B}+1395b_{1}
^{10}b_{2}^{4}\varepsilon\sin\mathbb{B}\cos^{5}\mathbb{B}-900b_{1}^{8}
b_{2}^{6}\varepsilon\sin\mathbb{B}\cos^{5}\mathbb{B}
\\\nonumber
&+&882b_{1}^{4}b_{2}^{10}\varepsilon\sin\mathbb{B}\cos^{5}\mathbb{B}-1575
b_{1}^{6}b_{2}^{8}\varepsilon\sin\mathbb{B}\cos^{5}\mathbb{B}-63b_{2}^{12}
b_{1}^{2}\varepsilon\sin\mathbb{B}\cos^{5}\mathbb{B}
\\\nonumber
&+&195b_{2}^{12}b_{1}^{2}\varepsilon\sin\mathbb{B}\cos^{11}\mathbb{B}-1287
b_{1}^{4}b_{2}^{10}\varepsilon\sin\mathbb{B}\cos^{11}\mathbb{B}-3312
\varepsilon^{2}\sin\mathbb{B}\cos^{9}\mathbb{B}
\\\nonumber
&\times&b_{1}^{12}b_{2}^{2}\-17820b_{1}^{8}b_{2}^{6}\varepsilon^{2}\sin
\mathbb{B}\cos^{9}\mathbb{B}+20394b_{1}^{10}b_{2}^{4}\varepsilon^{2}
\sin\mathbb{B}\cos^{9}\mathbb{B}-20790\cos^{9}\mathbb{B}
\\\nonumber
&\times&b_{1}^{6}b_{2}^{8}\varepsilon^{2}\sin\mathbb{B}-2538b_{1}^{2}
b_{2}^{12}\varepsilon^{2}\sin\mathbb{B}\cos^{9}\mathbb{B}+18216b_{1}
^{4}b_{2}^{10}\varepsilon^{2}\sin\mathbb{B}\cos^{9}\mathbb{B}-552
\\\nonumber
&\times&b_{1}^{12}b_{2}^{2}\varepsilon\sin\mathbb{B}\cos^{9}\mathbb{B}
+3399b_{1}^{10}b_{2}^{4}\varepsilon\sin\mathbb{B}\cos^{9}\mathbb{B}
-2970b_{1}^{8}b_{2}^{6}\varepsilon\sin\mathbb{B}\cos^{9}\mathbb{B}
-b_{1}^{6}
\\\nonumber
&\times&3465b_{2}^{8}\varepsilon\sin\mathbb{B}\cos^{9}\mathbb{B}+3036
b_{1}^{4}b_{2}^{10}\varepsilon\sin\mathbb{B}\cos^{9}\mathbb{B}-423b_{2}
^{12}b_{1}^{2}\varepsilon\sin\mathbb{B}\cos^{9}\mathbb{B}+\varepsilon^{2}
\\\nonumber
&\times&+3294b_{1}^{12}b_{2}^{2}\sin\mathbb{B}\cos^{7}\mathbb{B}-19602b_{1}
^{10}b_{2}^{4}\varepsilon^{2}\sin\mathbb{B}\cos^{7}\mathbb{B}+14850
\varepsilon^{2}\sin\mathbb{B}\cos^{7}\mathbb{B}
\\\nonumber
&\times&b_{1}^{8}b_{2}^{6}+1746b_{1}^{2}b_{2}^{12}\varepsilon^{2}\sin\mathbb{B}
\cos^{7}\mathbb{B}+20790b_{1}^{6}b_{2}^{8}\varepsilon^{2}\sin\mathbb{B}\cos^{7}
\mathbb{B}-15246b_{1}^{4}\cos^{7}\mathbb{B}
\\\nonumber
&\times&b_{2}^{10}\varepsilon^{2}\sin\mathbb{B}+810b_{1}^{8}b_{2}^{6}
\varepsilon^{2}\cos^{3}\mathbb{B}\sin\mathbb{B}+180b_{1}^{12}b_{2}^{2}
\varepsilon^{2}\sin\mathbb{B}\cos^{3}\mathbb{B}-1530\cos^{3}\mathbb{B}
\\\nonumber
&\times&b_{1}^{10}b_{2}^{4}\varepsilon^{2}\sin\mathbb{B}+1170b_{1}^{12}b_{2}^{2}
\varepsilon^{2}\sin\mathbb{B}\cos^{11}\mathbb{B}-7722b_{1}^{10}b_{2}^{4}
\varepsilon^{2}\sin\mathbb{B}\cos^{11}\mathbb{B}+\sin\mathbb{B}
\\\nonumber
&\times&7722b_{1}^{8}b_{2}^{6}\varepsilon^{2}\cos^{11}\mathbb{B}+1890b_{1}^{6}
b_{2}^{8}\varepsilon^{2}\sin\mathbb{B}\cos^{3}\mathbb{B}-630b_{1}^{4}b_{2}^{10}
\varepsilon^{2}\sin\mathbb{B}\cos^{3}\mathbb{B}+b_{1}^{6}
\\\nonumber
&\times&7722b_{2}^{8}\varepsilon^{2}\sin\mathbb{B}\cos^{11}\mathbb{B}
-7722b_{1}^{4}b_{2}^{10}\varepsilon^{2}\sin\mathbb{B}\cos^{11}\mathbb{B}
+1170b_{1}^{2}\varepsilon^{2}\sin\mathbb{B}\cos^{11}\mathbb{B}
\\\nonumber
&\times&b_{2}^{12}+195b_{2}^{2}\varepsilon\sin\mathbb{B}b_{1}^{12}
\cos^{11}\mathbb{B}-1287b_{1}^{10}b_{2}^{4}\varepsilon\sin\mathbb{B}
\cos^{11}\mathbb{B}+1287b_{1}^{8}b_{2}^{6}\cos^{11}\mathbb{B}
\\\nonumber
&\times&\varepsilon\sin\mathbb{B}+1287b_{1}^{6}b_{2}^{8}\varepsilon
\sin\mathbb{B}\cos^{11}\mathbb{B}+90b_{1}^{10}b_{2}^{4}\varepsilon^{2}
\sin\mathbb{B}\cos\mathbb{B}-36\varepsilon^{2}\sin\mathbb{B}\cos\mathbb{B}
\\\nonumber
&\times&b_{1}^{8}b_{2}^{6}-126b_{1}^{6}b_{2}^{8}\varepsilon^{2}\sin\mathbb{B}
\cos\mathbb{B}+15b_{1}^{10}b_{2}^{4}\varepsilon\sin\mathbb{B}\cos\mathbb{B}
-6b_{1}^{8}b_{2}^{6}\varepsilon\sin\mathbb{B}\cos\mathbb{B}-b_{1}^{6}
\\\nonumber
&\times&21b_{2}^{8}\varepsilon\sin\mathbb{B}\cos\mathbb{B}+b_{1}^{14}\sin
\mathbb{B}\cos^{11}\mathbb{B}+b_{2}^{14}\sin\mathbb{B}\cos^{11}\mathbb{B}
+180b_{1}^{11}b_{2}^{3}\varepsilon^{2}\cos^{2}\mathbb{B}
\\\nonumber
&+&378b_{1}^{5}b_{2}^{9}\varepsilon^{2}\cos^{2}\mathbb{B}+30b_{1}^{11}b_{2}
^{3}\varepsilon\cos^{2}\mathbb{B}-558b_{1}^{9}b_{2}^{5}\varepsilon^{2}\cos^{2}
\mathbb{B}-360b_{1}^{7}b_{2}^{7}\varepsilon^{2}\cos^{2}\mathbb{B}
\\\nonumber
&+&b_{2}^{14}\sin\mathbb{B}\cos^{7}\mathbb{B}-b_{1}^{14}\sin\mathbb{B}
\cos^{5}\mathbb{B}-93b_{1}^{9}b_{2}^{5}\varepsilon\cos^{2}\mathbb{B}
-60b_{1}^{7}b_{2}^{7}\varepsilon\cos^{2}\mathbb{B}-\sin\mathbb{B}
\\\nonumber
&\times&5b_{1}^{10}b_{2}^{4}\cos\mathbb{B}+63b_{1}^{5}b_{2}^{9}\varepsilon
\cos^{2}\mathbb{B}+2b_{1}^{8}b_{2}^{6}\sin\mathbb{B}\cos\mathbb{B}
+7b_{1}^{6}b_{2}^{8}\sin\mathbb{B}\cos\mathbb{B}-\cos^{4}\mathbb{B}
\\\nonumber
&-&4410b_{1}^{5}b_{2}^{9}\varepsilon^{2}+630b_{1}^{3}b_{2}^{11}
\varepsilon^{2}\cos^{4}\mathbb{B}+15b_{1}^{13}b_{2}\varepsilon
\cos^{4}\mathbb{B}-345b_{1}^{11}b_{2}^{3}\varepsilon\cos^{4}
\mathbb{B}+\cos^{4}\mathbb{B}
\\\nonumber
&\times&840b_{1}^{9}b_{2}^{5}\varepsilon+360b_{1}^{7}b_{2}^{7}\varepsilon
\cos^{4}\mathbb{B}-735b_{1}^{5}b_{2}^{9}\varepsilon\cos^{4}\mathbb{B}
+105b_{1}^{3}b_{2}^{11}\varepsilon\cos^{4}\mathbb{B}-10\cos^{3}\mathbb{B}
\\\nonumber
&\times&b_{1}^{12}b_{2}^{2}\sin\mathbb{B}+85b_{1}^{10}b_{2}^{4}\sin
\mathbb{B}\cos^{3}\mathbb{B}-45b_{1}^{8}b_{2}^{6}\sin\mathbb{B}
\cos^{3}\mathbb{B}-105b_{1}^{6}b_{2}^{8}\sin\mathbb{B}\cos^{3}\mathbb{B}
\\\nonumber
&+&35b_{1}^{4}b_{2}^{10}\sin\mathbb{B}\cos^{3}\mathbb{B}-78b_{2}^{13}b_{1}
\varepsilon\cos^{8}\mathbb{B}-3b_{2}^{14}\varepsilon\sin\mathbb{B}\cos^{7}
\mathbb{B}-183b_{1}^{12}b_{2}^{2}\cos^{7}\mathbb{B}
\\\nonumber
&\times&\sin\mathbb{B}-9b_{1}^{14}\varepsilon\sin\mathbb{B}\cos^{7}\mathbb{B}
-825b_{1}^{8}b_{2}^{6}\sin\mathbb{B}\cos^{7}\mathbb{B}-1155b_{1}^{6}b_{2}^{8}
\sin\mathbb{B}\cos^{7}\mathbb{B}+b_{1}^{10}
\\\nonumber
&\times&1089b_{2}^{4}\sin\mathbb{B}\cos^{7}\mathbb{B}-97b_{2}^{12}b_{1}^{2}
\sin\mathbb{B}\cos^{7}\mathbb{B}-486b_{1}^{13}b_{2}\varepsilon^{2}\cos^{6}
\mathbb{B}+847b_{1}^{4}\cos^{7}\mathbb{B}
\\\nonumber
&\times&b_{2}^{10}\sin\mathbb{B}+8244b_{1}^{11}b_{2}^{3}\varepsilon^{2}
\cos^{6}\mathbb{B}-19350b_{1}^{9}b_{2}^{5}\varepsilon^{2}\cos^{6}\mathbb{B}
-5400b_{1}^{7}b_{2}^{7}\varepsilon^{2}\cos^{6}\mathbb{B}+b_{1}^{5}
\\\nonumber
&\times&18270b_{2}^{9}\varepsilon^{2}\cos^{6}\mathbb{B}-4284b_{1}^{3}b_{2}
^{11}\varepsilon^{2}\cos^{6}\mathbb{B}+126b_{1}b_{2}^{13}\varepsilon^{2}
\cos^{6}\mathbb{B}+18\varepsilon^{2}\sin\mathbb{B}\cos^{5}\mathbb{B}
\\\nonumber
&\times&b_{1}^{14}+1374b_{1}^{11}b_{2}^{3}\varepsilon\cos^{6}\mathbb{B}
-3225b_{1}^{9}b_{2}^{5}\varepsilon\cos^{6}\mathbb{B}-900b_{1}^{7}b_{2}^{7}
\varepsilon\cos^{6}\mathbb{B}+3045b_{1}^{5}\varepsilon\cos^{6}\mathbb{B}
\\\nonumber
&\times&b_{2}^{9}-714b_{1}^{3}b_{2}^{11}\varepsilon\cos^{6}\mathbb{B}
+21b_{1}b_{2}^{13}\varepsilon\cos^{6}\mathbb{B}+3b_{1}^{14}\varepsilon
\sin\mathbb{B}\cos^{5}\mathbb{B}+74b_{1}^{12}\sin\mathbb{B}\cos^{5}
\mathbb{B}
\\\nonumber
&\times&b_{2}^{2}-465b_{1}^{10}b_{2}^{4}\sin\mathbb{B}\cos^{5}\mathbb{B}
+300b_{1}^{8}b_{2}^{6}\sin\mathbb{B}\cos^{5}\mathbb{B}+525b_{1}^{6}b_{2}
^{8}\sin\mathbb{B}\cos^{5}\mathbb{B}-\cos^{6}\mathbb{B}
\\\nonumber
&\times&81b_{1}^{13}b_{2}\varepsilon-294b_{1}^{4}b_{2}^{10}\sin\mathbb{B}
\cos^{5}\mathbb{B}+21b_{2}^{12}b_{1}^{2}\sin\mathbb{B}\cos^{5}\mathbb{B}
+90b_{1}^{13}b_{2}\varepsilon^{2}\cos^{4}\mathbb{B}-\varepsilon^{2}b_{1}^{11}
\\\nonumber
&\times&2070b_{2}^{3}\cos^{4}\mathbb{B}+5040b_{1}^{9}b_{2}^{5}\varepsilon^{2}
\cos^{4}\mathbb{B}+2160b_{1}^{7}b_{2}^{7}\varepsilon^{2}\cos^{4}\mathbb{B}
+216b_{1}^{13}b_{2}\varepsilon^{2}\cos^{12}\mathbb{B}-b_{1}^{11}
\\\nonumber
&\times&3744b_{2}^{3}\varepsilon^{2}\cos^{12}\mathbb{B}+10296b_{1}^{9}b_{2}^{5}
\varepsilon^{2}\cos^{12}\mathbb{B}-10296b_{1}^{5}b_{2}^{9}\varepsilon^{2}\cos^{12}
\mathbb{B}-216b_{1}\varepsilon^{2}\cos^{12}\mathbb{B}
\\\nonumber
&\times&b_{2}^{13}+3744b_{1}^{3}b_{2}^{11}\varepsilon^{2}\cos^{12}\mathbb{B}
-18b_{1}^{14}\varepsilon^{2}\sin\mathbb{B}\cos^{11}\mathbb{B}+36b_{1}^{13}b_{2}
\varepsilon\cos^{12}\mathbb{B}-624\cos^{12}\mathbb{B}
\\\nonumber
&\times&b_{1}^{11}b_{2}^{3}\varepsilon-18b_{2}^{14}\varepsilon^{2}\sin\mathbb{B}
\cos^{11}\mathbb{B}+1716b_{1}^{9}b_{2}^{5}\varepsilon\cos^{12}\mathbb{B}
+624b_{1}^{3}b_{2}^{11}\varepsilon\cos^{12}\mathbb{B}-36\cos^{12}\mathbb{B}
\\\nonumber
&\times&b_{1}b_{2}^{13}\varepsilon-1716b_{1}^{5}b_{2}^{9}\varepsilon
\cos^{12}\mathbb{B}-3b_{2}^{14}\varepsilon\sin\mathbb{B}\cos^{11}\mathbb{B}
-65b_{1}^{12}b_{2}^{2}\sin\mathbb{B}\cos^{11}\mathbb{B}-3\cos^{11}\mathbb{B}
\\\nonumber
&\times&\varepsilon
b_{1}^{14}\sin\mathbb{B}+429b_{1}^{10}b_{2}^{4}\sin\mathbb{B}\cos^{11}\mathbb{B}
-429b_{1}^{8}b_{2}^{6}\sin\mathbb{B}\cos^{11}\mathbb{B}-429b_{1}^{6}b_{2}^{8}
\sin\mathbb{B}\cos^{11}\mathbb{B}
\\\nonumber
&+&429b_{1}^{4}b_{2}^{10}\sin\mathbb{B}\cos^{11}\mathbb{B}-65b_{1}^{2}b_{2}^{12}
\sin\mathbb{B}\cos^{11}\mathbb{B}-738b_{1}^{13}b_{2}\varepsilon^{2}\cos^{10}\mathbb{B}
-31086\cos^{10}\mathbb{B}
\\\nonumber
&\times&b_{1}^{9}b_{2}^{5}\varepsilon^{2}+12132b_{1}^{11}b_{2}^{3}\varepsilon^{2}
\cos^{10}\mathbb{B}-2376b_{1}^{7}b_{2}^{7}\varepsilon^{2}\cos^{10}\mathbb{B}
-10332b_{1}^{3}b_{2}^{11}\varepsilon^{2}\cos^{10}\mathbb{B}+b_{1}^{5}b_{2}^{9}
\\\nonumber
&\times&30690\varepsilon^{2}\cos^{10}\mathbb{B}+558b_{1}b_{2}^{13}\varepsilon^{2}
\cos^{10}\mathbb{B}+54b_{1}^{14}\varepsilon^{2}\sin\mathbb{B}\cos^{9}\mathbb{B}
+36\sin\mathbb{B}\cos^{9}\mathbb{B}
\\\nonumber
&\times&b_{2}^{14}\varepsilon^{2}+2022b_{1}^{11}b_{2}^{3}\varepsilon\cos^{10}\mathbb{B}
-123b_{1}^{13}b_{2}\varepsilon\cos^{10}\mathbb{B}-5181b_{1}^{9}b_{2}^{5}\varepsilon
\cos^{10}\mathbb{B}-\cos^{10}\mathbb{B}
\\\nonumber
&\times&396b_{1}^{7}b_{2}^{7}\varepsilon-1722b_{1}^{3}b_{2}^{11}\varepsilon\cos^{10}
\mathbb{B}+5115b_{1}^{5}b_{2}^{9}\cos^{10}\mathbb{B}\varepsilon+93b_{1}b_{2}^{13}
\varepsilon\cos^{10}\mathbb{B}+\sin\mathbb{B}
\\\nonumber
&\times&9\varepsilon
b_{1}^{14}\cos^{9}\mathbb{B}+184b_{1}^{12}b_{2}^{2}\sin\mathbb{B}\cos^{9}\mathbb{B}
-1133b_{1}^{10}b_{2}^{4}\sin\mathbb{B}\cos^{9}\mathbb{B}+990b_{1}^{8}b_{2}^{6}
\cos^{9}\mathbb{B}
\\\nonumber
&\times&\sin\mathbb{B}+1155b_{1}^{6}b_{2}^{8}\sin\mathbb{B}\cos^{9}\mathbb{B}
-1012b_{1}^{4}b_{2}^{10}\sin\mathbb{B}\cos^{9}\mathbb{B}+141b_{2}^{12}b_{1}^{2}
\sin\mathbb{B}\cos^{9}\mathbb{B}
\\\nonumber
&+&918b_{1}^{13}b_{2}\varepsilon^{2}\cos^{8}\mathbb{B}-14742b_{1}^{11}b_{2}^{3}
\varepsilon^{2}\cos^{8}\mathbb{B}+35640b_{1}^{9}b_{2}^{5}\varepsilon^{2}\cos^{8}
\mathbb{B}+5940\cos^{8}\mathbb{B}
\\\nonumber
&\times&b_{1}^{7}b_{2}^{7}\varepsilon^{2}-34650b_{1}^{5}b_{2}^{9}\varepsilon^{2}
\cos^{8}\mathbb{B}+10242b_{1}^{3}b_{2}^{11}\varepsilon^{2}\cos^{8}\mathbb{B}
-468b_{1}b_{2}^{13}\varepsilon^{2}\cos^{8}\mathbb{B}-b_{1}^{14}
\\\nonumber
&\times&54\varepsilon^{2}\sin\mathbb{B}\cos^{7}\mathbb{B}-18b_{2}^{14}
\varepsilon^{2}\sin\mathbb{B}\cos^{7}\mathbb{B}+153b_{1}^{13}b_{2}
\varepsilon\cos^{8}\mathbb{B}-2457b_{1}^{11}\cos^{8}\mathbb{B}
\\\nonumber
&\times&\varepsilon
b_{2}^{3}+990b_{1}^{7}b_{2}^{7}\varepsilon\cos^{8}
\mathbb{B}-5775b_{1}^{5}b_{2}^{9}\varepsilon\cos^{8}\mathbb{B}
+1707b_{1}^{3}b_{2}^{11}\varepsilon\cos^{8}\mathbb{B}+3b_{1}^{7}
b_{2}^{7}\varepsilon+b_{1}^{9}
\\\nonumber
&\times&18b_{2}^{5}\varepsilon^{2}+18b_{1}^{7}b_{2}^{7}\varepsilon^{2}
+5940b_{1}^{9}b_{2}^{5}\varepsilon\cos^{8}\mathbb{B}+6\varepsilon\sin
\mathbb{B}b_{2}^{14}\cos^{9}\mathbb{B}-12b_{1}^{13}b_{2}\cos^{12}\mathbb{B}
\\\nonumber
&+&208b_{1}^{11}b_{2}^{3}\cos^{12}\mathbb{B}-572b_{1}^{9}b_{2}^{5}\cos^{12}
\mathbb{B}+572b_{1}^{5}b_{2}^{9}\cos^{12}\mathbb{B}+12b_{1}b_{2}^{13}
\cos^{12}\mathbb{B}-b_{1}^{3}
\\\nonumber
&\times&208b_{2}^{11}\cos^{12}\mathbb{B}+41b_{1}^{13}b_{2}\cos^{10}\mathbb{B}
-674b_{1}^{11}b_{2}^{3}\cos^{10}\mathbb{B}+132b_{1}^{7}b_{2}^{7}
\cos\mathbb{B}^{10}-b_{1}^{5}
\\\nonumber
&\times&1705b_{2}^{9}\cos^{10}\mathbb{B}+1727b_{1}^{9}b_{2}^{5}\cos^{10}\mathbb{B}
+574b_{1}^{3}b_{2}^{11}\cos^{10}\mathbb{B}-31b_{1}b_{2}^{13}\cos^{10}\mathbb{B}-3
\\\nonumber
&\times&b_{1}^{14}\sin\mathbb{B}\cos^{9}\mathbb{B}-51b_{1}^{13}b_{2}
\cos^{8}\mathbb{B}-2\sin\mathbb{B}b_{2}^{14}\cos^{9}\mathbb{B}
-1980b_{1}^{9}b_{2}^{5}\cos^{8}\mathbb{B}+b_{1}^{11}
\\\nonumber
&\times&819b_{2}^{3}\cos^{8}\mathbb{B}-330b_{1}^{7}b_{2}^{7}\cos^{8}\mathbb{B}
+1925b_{1}^{5}b_{2}^{9}\cos^{8}\mathbb{B}+3b_{1}^{14}\sin\mathbb{B}\cos^{7}
\mathbb{B}-b_{1}^{3}b_{2}^{11}
\\\nonumber
&\times&569\cos^{8}\mathbb{B}+26b_{1}b_{2}^{13}\cos^{8}\mathbb{B}+27b_{1}^{13}
b_{2}\cos^{6}\mathbb{B}+1075b_{1}^{9}b_{2}^{5}\cos^{6}\mathbb{B}+\cos^{6}
\mathbb{B}b_{1}^{7}
\\\nonumber
&\times&300b_{2}^{7}-458b_{1}^{11}b_{2}^{3}\cos^{6}\mathbb{B}-1015b_{1}^{5}b_{2}
^{9}\cos^{6}\mathbb{B}-7b_{1}b_{2}^{13}\cos^{6}\mathbb{B}-5b_{1}^{13}b_{2}
\cos^{4}\mathbb{B}
\\\nonumber
&+&238b_{1}^{3}b_{2}^{11}\cos^{6}\mathbb{B}+115b_{1}^{11}b_{2}^{3}\cos^{4}\mathbb{B}
-120b_{1}^{7}b_{2}^{7}\cos^{4}\mathbb{B}+245b_{1}^{5}b_{2}^{9}\cos^{4}\mathbb{B}
-b_{1}^{9}b_{2}^{5}
\\\nonumber
&\times&280\cos^{4}\mathbb{B}-35b_{1}^{3}b_{2}^{11}\cos^{4}\mathbb{B}
+31b_{1}^{9}b_{2}^{5}\cos^{2}\mathbb{B}+20b_{1}^{7}b_{2}^{7}\cos^{2}
\mathbb{B}-21b_{1}^{5}b_{2}^{9}\cos^{2}\mathbb{B}
\\\nonumber
&-&10b_{1}^{11}b_{2}^{3}\cos^{2}\mathbb{B}+3b_{1}^{9}b_{2}^{5}\varepsilon
-b_{1}^{9}b_{2}^{5}-b_{1}^{7}b_{2}^{7}\bigg]\bigg[-60b_{1}^{12}b_{2}^{2}
\varepsilon\sin\mathbb{B}\cos^{3}\mathbb{B}+510\cos^{3}\mathbb{B}
\\\nonumber
&\times&b_{1}^{10}b_{2}^{4}\varepsilon\sin\mathbb{B}-630b_{1}^{6}b_{2}^{8}
\varepsilon\sin\mathbb{B}\cos^{3}\mathbb{B}+210b_{1}^{4}b_{2}^{10}
\varepsilon\sin\mathbb{B}\cos^{3}\mathbb{B}-1098b_{1}^{12}\cos^{7}\mathbb{B}
\\\nonumber
&\times&b_{2}^{2}\varepsilon\sin\mathbb{B}+6534b_{1}^{10}b_{2}^{4}\varepsilon
\sin\mathbb{B}\cos^{7}\mathbb{B}-4950b_{1}^{8}b_{2}^{6}\varepsilon\sin
\mathbb{B}\cos^{7}\mathbb{B}-6930b_{1}^{6}\cos^{7}\mathbb{B}
\\\nonumber
&\times&b_{2}^{8}\varepsilon\sin\mathbb{B}+5082b_{1}^{4}b_{2}^{10}
\varepsilon\sin\mathbb{B}\cos^{7}\mathbb{B}-582b_{2}^{12}b_{1}^{2}
\varepsilon\sin\mathbb{B}\cos^{7}\mathbb{B}+444b_{1}^{12}\cos^{5}
\mathbb{B}
\\\nonumber
&\times&b_{2}^{2}\varepsilon\sin\mathbb{B}-72790b_{1}^{10}b_{2}^{4}
\varepsilon\sin\mathbb{B}\cos^{5}\mathbb{B}+1800b_{1}^{8}b_{2}^{6}
\varepsilon\sin\mathbb{B}\cos^{5}\mathbb{B}+3150\cos^{5}\mathbb{B}
\\\nonumber
&\times&b_{1}^{6}b_{2}^{8}\varepsilon\sin\mathbb{B}-1764b_{1}^{4}
b_{2}^{10}\varepsilon\sin\mathbb{B}\cos^{5}\mathbb{B}+126b_{2}
^{12}b_{1}^{2}\varepsilon\sin\mathbb{B}\cos^{5}\mathbb{B}+2574
\cos^{11}\mathbb{B}
\\\nonumber
&\times&b_{1}^{4}b_{2}^{10}\varepsilon\sin\mathbb{B}-390b_{2}^{12}
b_{1}^{2}\varepsilon\sin\mathbb{B}\cos^{11}\mathbb{B}+1104b_{1}
^{12}b_{2}^{2}\varepsilon\sin\mathbb{B}\cos^{9}\mathbb{B}-6798
b_{1}^{10}b_{2}^{4}
\\\nonumber
&\times&\varepsilon\sin\mathbb{B}\cos^{9}\mathbb{B}+5940b_{1}
^{8}b_{2}^{6}\varepsilon\sin\mathbb{B}\cos^{9}\mathbb{B}+6930
b_{1}^{6}b_{2}^{8}\varepsilon\sin\mathbb{B}\cos^{9}\mathbb{B}
-6072b_{1}^{4}b_{2}^{10}
\\\nonumber
&\times&\varepsilon\sin\mathbb{B}\cos^{9}\mathbb{B}+846b_{1}^{2}
\varepsilon\sin\mathbb{B}b_{2}^{12}\cos^{9}\mathbb{B}-390b_{2}^{2}
\varepsilon\sin\mathbb{B}b_{1}^{12}\cos^{11}\mathbb{B}+2574b_{1}
^{10}b_{2}^{4}
\\\nonumber
&\times&\varepsilon\sin\mathbb{B}\cos^{11}\mathbb{B}-2574b_{1}^{8}
b_{2}^{6}\varepsilon\sin\mathbb{B}\cos^{11}\mathbb{B}-2574b_{1}^{6}
b_{2}^{8}\varepsilon\sin\mathbb{B}\cos^{11}\mathbb{B}-30b_{1}^{10}
b_{2}^{4}
\\\nonumber
&\times&\varepsilon\sin\mathbb{B}\cos\mathbb{B}+12b_{1}^{8}b_{2}^{6}
\varepsilon\sin\mathbb{B}\cos\mathbb{B}+42b_{1}^{6}b_{2}^{8}
\varepsilon\sin\mathbb{B}\cos\mathbb{B}+11b_{1}^{14}\sin\mathbb{B}
\cos^{11}\mathbb{B}
\\\nonumber
&-&270b_{1}^{8}b_{2}^{6}\varepsilon\sin\mathbb{B}\cos^{3}\mathbb{B}
-7b_{2}^{14}\sin\mathbb{B}\cos^{11}\mathbb{B}-b_{2}^{14}\sin
\mathbb{B}\cos^{7}\mathbb{B}-2b_{1}^{14}\sin\mathbb{B}\cos^{5}
\mathbb{B}
\\\nonumber
&+&186b_{1}^{9}b_{2}^{5}\varepsilon\cos^{2}\mathbb{B}-60b_{1}^{11}
b_{2}^{3}\varepsilon\cos^{2}\mathbb{B}+120b_{1}^{7}b_{2}^{7}
\varepsilon\cos^{2}\mathbb{B}-126b_{1}^{5}b_{2}^{9}\varepsilon
\cos^{2}\mathbb{B}-10b_{1}^{10}
\\\nonumber
&\times&b_{2}^{4}\sin\mathbb{B}\cos\mathbb{B}+25b_{1}^{8}b_{2}^{6}
\sin\mathbb{B}\cos\mathbb{B}-7b_{1}^{6}b_{2}^{8}\sin\mathbb{B}
\cos\mathbb{B}-30b_{1}^{13}b_{2}\varepsilon\cos^{4}\mathbb{B}+690
\\\nonumber
&\times&b_{1}^{11}b_{2}^{3}\varepsilon\cos^{4}\mathbb{B}-1680b_{1}^{9}
b_{2}^{5}\varepsilon\cos^{4}\mathbb{B}-720b_{1}^{7}b_{2}^{7}\varepsilon
\cos^{4}\mathbb{B}+1470b_{1}^{5}b_{2}^{9}\varepsilon\cos^{4}\mathbb{B}-210
\\\nonumber
&\times&b_{1}^{3}b_{2}^{11}\varepsilon\cos^{4}\mathbb{B}-20b_{1}^{12}
b_{2}^{2}\sin\mathbb{B}\cos^{3}\mathbb{B}+275b_{1}^{10}b_{2}^{4}
\sin\mathbb{B}\cos^{3}\mathbb{B}-657\sin\mathbb{B}\cos^{3}\mathbb{B}
\\\nonumber
&\times&b_{1}^{8}b_{2}^{6}+357b_{1}^{6}b_{2}^{8}\sin\mathbb{B}\cos^{3}
\mathbb{B}-35b_{1}^{4}b_{2}^{10}\sin\mathbb{B}\cos^{3}\mathbb{B}+
156b_{1}b_{2}^{13}\varepsilon\cos^{8}\mathbb{B}+\cos^{7}\mathbb{B}
\\\nonumber
&\times&18b_{1}^{14}\varepsilon\sin\mathbb{B}+6b_{2}^{14}\varepsilon
\sin\mathbb{B}\cos^{7}\mathbb{B}-765b_{1}^{12}b_{2}^{2}\sin\mathbb{B}
\cos^{7}\mathbb{B}-18975\sin\mathbb{B}\cos^{7}\mathbb{B}
\\\nonumber
&\times&b_{1}^{8}b_{2}^{6}+7491b_{1}^{10}b_{2}^{4}\sin\mathbb{B}
\cos^{7}\mathbb{B}+15015b_{1}^{6}b_{2}^{8}\sin\mathbb{B}\cos^{7}
\mathbb{B}+205b_{2}^{12}b_{1}^{2}\sin\mathbb{B}\cos^{7}\mathbb{B}
\\\nonumber
&-&3619b_{1}^{4}b_{2}^{10}\sin\mathbb{B}\cos^{7}\mathbb{B}+162b_{1}
^{13}b_{2}\varepsilon\cos^{6}\mathbb{B}+6450b_{1}^{9}b_{2}^{5}
\varepsilon\cos^{6}\mathbb{B}+1800\varepsilon\cos^{6}\mathbb{B}
\\\nonumber
&\times&b_{1}^{7}b_{2}^{7}-2748b_{1}^{11}b_{2}^{3}\varepsilon\cos^{6}
\mathbb{B}-6090b_{1}^{5}b_{2}^{9}\varepsilon\cos^{6}\mathbb{B}
-42b_{2}^{13}\varepsilon\cos^{6}\mathbb{B}b_{1}-6b_{1}^{14}\cos^{5}
\mathbb{B}
\\\nonumber
&\times&\varepsilon\sin\mathbb{B}+1428b_{1}^{3}b_{2}^{11}\varepsilon
\cos^{6}\mathbb{B}+211b_{1}^{12}b_{2}^{2}\sin\mathbb{B}\cos^{5}
\mathbb{B}+5325b_{1}^{8}b_{2}^{6}\sin\mathbb{B}\cos^{5}\mathbb{B}
\\\nonumber
&-&2190b_{1}^{10}b_{2}^{4}\sin\mathbb{B}\cos^{5}\mathbb{B}-3675b_{1}
^{6}b_{2}^{8}\sin\mathbb{B}\cos^{5}\mathbb{B}+672b_{1}^{4}b_{2}^{10}
\sin\mathbb{B}\cos^{5}\mathbb{B}-21
\\\nonumber
&\times&b_{2}^{12}b_{1}^{2}\sin\mathbb{B}\cos^{5}\mathbb{B}-72b_{1}
^{13}b_{2}\varepsilon\cos^{12}\mathbb{B}+1248b_{1}^{11}b_{2}^{3}
\varepsilon\cos^{12}\mathbb{B}-3432\varepsilon\cos^{12}\mathbb{B}
\\\nonumber
&\times&b_{1}^{9}b_{2}^{5}+72b_{2}^{13}b_{1}\varepsilon\cos^{12}\mathbb{B}
+6b_{1}^{14}\varepsilon\sin\mathbb{B}\cos^{11}\mathbb{B}+6b_{2}{14}
\varepsilon\sin\mathbb{B}\cos^{11}\mathbb{B}-b_{1}^{12}b_{2}^{2}
\\\nonumber
&\times&949\sin\mathbb{B}\cos^{11}\mathbb{B}+9867b_{1}^{10}b_{2}^{4}
\sin\mathbb{B}\cos(\mathbb{B})^{11}-27885b_{1}^{8}b_{2}^{6}\sin\mathbb{B}
\cos^{11}\mathbb{B}+b_{1}^{5}
\\\nonumber
&\times&3432b_{2}^{9}\varepsilon\cos^{12}\mathbb{B}+26169b_{1}^{6}b_{2}^{8}
\sin\mathbb{B}\cos^{11}\mathbb{B}-8151b_{1}^{4}b_{2}^{10}\sin\mathbb{B}
\cos^{11}\mathbb{B}+b_{1}^{13}b_{2}
\\\nonumber
&\times&246\cos^{10}\mathbb{B}\varepsilon-4044b_{1}^{11}b_{2}^{3}\cos^{10}
\mathbb{B}\varepsilon+10362b_{1}^{9}b_{2}^{5}\cos\mathbb{B}^{10}\varepsilon
+792b_{1}^{7}b_{2}^{7}\varepsilon\cos^{10}\mathbb{B}
\\\nonumber
&-&10230
b_{1}^{5}b_{2}^{9}\cos^{10}\mathbb{B}\varepsilon+3444b_{1}^{3}b_{2}^{11}
\cos\mathbb{B}^{10}\varepsilon-186b_{2}^{13}\cos^{10}\mathbb{B}b_{1}
\varepsilon-18b_{1}^{14}\sin\mathbb{B}
\\\nonumber
&\times&\cos\mathbb{B}^{9}\varepsilon-12\sin\mathbb{B}b_{2}^{14}\cos^{9}
\mathbb{B}\varepsilon+1250\cos\mathbb{B}^{9}\sin\mathbb{B}b_{1}^{12}b_{2}
^{2}-12430\cos^{9}\mathbb{B}\sin\mathbb{B}
\\\nonumber
&\times&b_{1}^{10}b_{2}^{4}+33165\cos^{9}\mathbb{B}\sin\mathbb{B}b_{1}^{8}
b_{2}^{6}-28875\sin\mathbb{B}b_{1}^{6}b_{2}^{8}\cos^{9}\mathbb{B}+8140
\cos^{9}\mathbb{B}\sin\mathbb{B}
\\\nonumber
&\times&b_{1}^{4}b_{2}^{10}-600\cos^{9}\mathbb{B}\sin\mathbb{B}b_{2}^{12}
b_{1}^{2}-306b_{1}^{13}\cos^{8}\mathbb{B}b_{2}\varepsilon+4914b_{1}^{11}
b_{2}^{3}\cos^{8}\mathbb{B}\varepsilon-b_{1}^{7}b_{2}^{7}
\\\nonumber
&\times&1980\varepsilon\cos^{8}\mathbb{B}+11550b_{1}^{5}b_{2}^{9}
\varepsilon\cos^{8}\mathbb{B}-3414b_{1}^{3}b_{2}^{11}\varepsilon
\cos^{8}\mathbb{B}-6b_{1}^{7}b_{2}^{7}\varepsilon+4238b_{1}^{11}
b_{2}^{3}
\\\nonumber
&\times&\cos^{12}\mathbb{B}-171b_{1}^{13}b_{2}\cos^{12}\mathbb{B}
-22165b_{1}^{9}b_{2}^{5}\cos^{12}\mathbb{B}-19877b_{1}^{5}b_{2}^{9}
\cos^{12}\mathbb{B}+3406b_{1}^{3}
\\\nonumber
&\times&b_{2}^{11}\cos^{12}\mathbb{B}-11880b_{1}^{9}b_{2}^{5}\cos^{8}
\mathbb{B}\varepsilon+689b_{2}^{12}b_{1}^{2}\sin\mathbb{B}\cos^{11}
\mathbb{B}-1248b_{1}^{3}b_{2}^{11}\varepsilon\cos^{12}\mathbb{B}
\\\nonumber
&-&123b_{2}^{13}b_{1}\cos^{12}\mathbb{B}+271b_{1}^{13}b_{2}\cos^{10}
\mathbb{B}-6472b_{1}^{11}b_{2}^{3}\cos^{10}\mathbb{B}+32329b_{1}^{9}
b_{2}^{5}\cos^{10}\mathbb{B}
\\\nonumber
&-&49632b_{1}^{7}b_{2}^{7}\cos^{10}\mathbb{B}+25465b_{1}^{5}b_{2}^{9}
\cos^{10}\mathbb{B}-3976b_{1}^{3}b_{2}^{11}\cos^{10}\mathbb{B}
+127b_{1}b_{2}^{13}\cos^{10}\mathbb{B}
\\\nonumber
&-&15b_{1}^{14}\sin\mathbb{B}\cos^{9}\mathbb{B}+33990b_{1}^{7}b_{2}^{7}
\cos^{8}\mathbb{B}-15785b_{1}^{5}b_{2}^{9}\cos^{8}\mathbb{B}+2117b_{1}
^{3}b_{2}^{11}\cos^{8}\mathbb{B}
\\\nonumber
&-&23595b_{1}^{9}b_{2}^{5}\cos^{8}\mathbb{B}-207b_{1}^{13}b_{2}\cos^{8}
\mathbb{B}+4893b_{1}^{11}b_{2}^{3}\cos^{8}\mathbb{B}-1882b_{1}^{11}b_{2}
^{3}\cos^{6}\mathbb{B}
\\\nonumber
&-&53b_{1}b_{2}^{13}\cos^{8}\mathbb{B}+9\sin\mathbb{B}b_{1}^{14}\cos^{7}
\mathbb{B}+75b_{1}^{13}b_{2}\cos^{6}\mathbb{B}+5\sin\mathbb{B}b_{2}^{14}
\cos^{9}\mathbb{B}+b_{1}^{9}
\\\nonumber
&\times&8975b_{2}^{5}\cos^{6}\mathbb{B}-12000b_{1}^{7}b_{2}^{7}\cos^{6}
\mathbb{B}+4795b_{1}^{5}b_{2}^{9}\cos^{6}\mathbb{B}-490b_{1}^{3}b_{2}^{11}
\cos^{6}\mathbb{B}+b_{2}^{13}
\\\nonumber
&\times&7b_{1}\cos^{6}\mathbb{B}-10b_{1}^{13}b_{2}\cos^{4}\mathbb{B}
+335b_{1}^{11}b_{2}^{3}\cos^{4}\mathbb{B}-1673b_{1}^{9}b_{2}^{5}\cos^{4}
\mathbb{B}+35b_{1}^{3}\cos^{4}\mathbb{B}
\\\nonumber
&\times&b_{2}^{11}+2028b_{1}^{7}b_{2}^{7}\cos^{4}\mathbb{B}-623b_{1}^{5}
b_{2}^{9}\cos^{4}\mathbb{B}-20b_{1}^{11}b_{2}^{3}\cos^{2}\mathbb{B}
-128b_{1}^{7}b_{2}^{7}\cos^{2}\mathbb{B}
\\\nonumber
&+&125b_{1}^{9}b_{2}^{5}\cos^{2}\mathbb{B}+21b_{1}^{5}b_{2}^{9}
\cos^{2}\mathbb{B}-6b_{1}^{9}b_{2}^{5}\varepsilon+42b_{1}^{13}b_{2}
\cos^{14}\mathbb{B}-3b_{1}^{14}\cos^{13}\mathbb{B}
\\\nonumber
&\times&\sin\mathbb{B}-1092b_{1}^{11}b_{2}^{3}\cos^{14}\mathbb{B}
+6006b_{1}^{9}b_{2}^{5}\cos^{14}\mathbb{B}-10296b_{1}^{7}b_{2}^{7}
\cos^{14}\mathbb{B}+6006
\\\nonumber
&\times&b_{1}^{5}b_{2}^{9}\cos^{14}\mathbb{B}-1092b_{1}^{3}b_{2}^{11}
\cos^{14}\mathbb{B}+42b_{1}b_{2}^{13}\cos^{14}\mathbb{B}-2b_{1}^{9}
b_{2}^{5}+b_{1}^{7}b_{2}^{7}-3003
\\\nonumber
&\times&b_{1}^{10}b_{2}^{4}\sin\mathbb{B}\cos^{13}\mathbb{B}+36036
b_{1}^{7}b_{2}^{7}\cos^{12}\mathbb{B}+273b_{1}^{12}b_{2}^{2}
\sin\mathbb{B}\cos^{13}\mathbb{B}+3\sin\mathbb{B}
\\\nonumber
&\times&b_{2}^{14}\cos^{13}\mathbb{B}+9009b_{1}^{8}b_{2}^{6}\sin\mathbb{B}
\cos^{13}\mathbb{B}-9009b_{1}^{6}b_{2}^{8}\sin\mathbb{B}\cos^{13}\mathbb{B}
+3003b_{1}^{4}b_{2}^{10}
\\\nonumber
&-&273b_{1}^{2}\sin\mathbb{B}b_{2}^{12}\cos^{13}\mathbb{B}\sin\mathbb{B}
\cos^{13}\mathbb{B}\bigg]^{-1}.
\end{eqnarray}\\
\textbf{Data availability:} No new data were generated or analyzed
in support of this research. \\\\
{\bf Acknowledgment}\\\\
This work has been supported by the \emph{Pakistan Academy of
Sciences Project}.

\end{document}